\documentclass{emulateapj}

\usepackage{lscape}
\usepackage{amsmath}
\usepackage{epstopdf}

\shortauthors{Constantin et al. }  

\shorttitle{X-ray LLAGN}

\begin{document}


\title{Probing the Balance of AGN and Star-Forming Activity in the
  Local Universe with ChaMP}

\author{Anca Constantin\altaffilmark{1}, 
  Paul Green\altaffilmark{1},
  Tom Aldcroft\altaffilmark{1},
  Dong-Woo Kim\altaffilmark{1}, 
  Daryl Haggard\altaffilmark{2}, 
  Wayne Barkhouse\altaffilmark{3}, 
  Scott F. Anderson\altaffilmark{2}}

\altaffiltext{1}{Harvard-Smithsonian Center for Astrophysics,
  Cambridge, MA 02138}
\altaffiltext{2}{Department of Astronomy, University of Washington,
  Seattle, WA, USA} 
\altaffiltext{3}{Department of Physics and Astrophysics, University of
  North Dakota, Grand Forks, ND 58202, USA}

\begin{abstract}

  The combination of the SDSS and the Chandra Multiwavelength Project
  (ChaMP) currently offers the largest and most homogeneously selected
  sample of nearby galaxies for investigating the relation between
  X-ray nuclear emission, nebular line-emission, black hole masses,
  and properties of the associated stellar populations.  We provide
  X-ray spectral fits and valid uncertainties for all the galaxies with counts
  ranging from 2 to 1325 (mean 76, median 19).  We present here
  novel constraints that both X-ray luminosity $L_X$ and X-ray
  spectral energy distribution bring to the galaxy evolutionary
  sequence {\it H II $\rightarrow$ Seyfert/Transition Object
    $\rightarrow$ LINER $\rightarrow$ Passive} suggested by optical
  data.  In particular, we show that both $L_X$ and $\Gamma$, the
  slope of the power-law that best fits the 0.5 - 8 keV spectra, are
  consistent with a clear decline in the accretion power along the
  sequence, corresponding to a softening of their spectra.  This
  implies that, at $z \approx 0$, or at low luminosity AGN levels,
  there is an anti-correlation between $\Gamma$ and $L/L_{\rm edd}$,
  opposite to the trend exhibited by high $z$ AGN (quasars).  The
  turning point in the $\Gamma -L/L_{\rm edd}$ LLAGN $+$ quasars
  relation occurs near $\Gamma \approx 1.5$ and $L/L_{\rm edd} \approx
  0.01$.  Interestingly, this is identical to what stellar mass X-ray
  binaries exhibit, indicating that we have probably found the first
  empirical evidence for an intrinsic switch in the accretion mode,
  from advection-dominated flows to standard (disk/corona) accretion
  modes in supermassive black hole accretors, similar to what has been
  seen and proposed to happen in stellar mass black hole systems.  The
  anti-correlation we find between $\Gamma$ and $L/L_{\rm edd}$ may
  instead indicate that stronger accretion correlates with greater
  absorption.  Therefore the trend for softer spectra toward more
  luminous, high redshift, and strongly accreting ($L/L_{edd} \ga
  0.01$) AGN/quasars could simply be the result of strong selection
  biases reflected in the dearth of type 2 quasar detections.
\end{abstract}

\keywords{galaxies: active -- galaxies: nuclei -- galaxies: emission
lines -- X-rays: galaxies -- surveys}

\section{Introduction: BH accretion, the low
  luminosity AGN, and X-ray emission}

Understanding the nuclear activity in nearby galaxies is essential for
constraining the galaxy formation and evolution process.  While
energetically unimpressive, the nearby galactic
nuclei offer the best case scenario for (1) the most common
state of accretion in the current universe, (2) the end-point of
quasar evolution, or simply (3) their scaled-down version.
Observationally, the low redshift $z \approx 0$ accretion systems are
the best testbeds for identifying the processes involved in triggering
and further fuelling accretion onto the central black hole (BH) because
they offer unique joint investigations of both the nuclear accretion
and the properties of the host, i.e., the star-formation (SF).

Unlike the optically luminous quasars, which are radiating close to
their Eddington limit \citep{kol06}, and only for short ($\sim 10^7$
yr) times \citep{yu02}, accretion activity in nearby galaxy centers
appears extremely diverse, spanning $> 6$ orders of magnitude in the
Eddington ratio, and (maybe consequently) a wide range in their
duty-cycles \citep{hec04, ho08}.  This variety provides empirical
constraints to model predictions linking the BH growth rate and the
host bulge formation.  The two are indeed connected: it is the younger
galaxies that host the more rapidly growing BHs
\citep{hec04,cidfernandes05,con08}.  The exact physical mechanism
responsible for this link, and in general for the close interplay
between SF and BH accretion remains however elusive.  Simulations
place important constraints on different models for the way black
holes are fueled, and provide a quantitative and physical distinction
between local, low luminosity, weak (or quiescent) AGN activity, and
violent, merger-driven bright quasars \citep{hopkins06, hopkins09}.
Environmental studies of the nearby active galactic nuclei are
consistent with these ideas of non-merger-driven fueling for the weak
BH growth observed in the nearby universe \citep{con06, con08}.  More
recent analysis of the observed distribution of Eddington ratios as a
function of the BH masses provide additional constraints, suggesting
that even at $z \approx 0$ there might be two distinct regimes of BH
growth, which are determined by the supply of cold gas in the host
bulge.  The BH regulates its own growth at a rate that is independent
of the interstellar medium's characteristics as long as the gas is
plentiful, but when the gas runs out the BH's growth will be regulated
by the rate at which evolved stars lose their mass
\citep{kauffmann09}.  These different fueling modes at low
luminosities must manifest differently at wavelengths outside the
optical regime, allowing further means to constrain
and discriminate among them.

The most ubiquitous type of activity at $z \approx 0$ that resembles
that of quasars is identified at optical wavelengths as either a
narrow-lined Low Ionization Nuclear Emission Regions(LINER, L) or a
``transition'' (T) object, whose properties border on the definition
of a starburst galaxy and an AGN.  Emission-line ratio diagnostics
\citep{bpt, vei87, kew06} that have been quite successful in
identifying cases where the dominant ionization mechanism is either
accretion onto a black hole (i.e., Seyferts, Ss) or radiation from
hot, young stars (i.e., H {\sc ii} nuclei), remain inconclusive for
the majority of Ls and Ts.  The Ls that exhibit quasar-like broad
lines (L1s, by analogy with Seyfert 1s) are unambiguously
accretion-powered sources, however those lacking these features (the
L2s) could have lines generated by shocks, post-starbursts, or other
processes unrelated with accretion.  Deciphering the
underlying emission source(s) of these ambiguous nuclei is an ongoing
struggle.

Recent analyses of the emission properties of the low luminosity AGN
(LLAGN) in relation to a wide variety of characteristics of their
hosts, together with considerations of their small and large scale
environments, reveal a sequence {\it H II $\rightarrow$
  Seyfert/Transition Object $\rightarrow$ LINER $\rightarrow$ Passive}
({\it H II $\rightarrow$ S/T $\rightarrow$ L $\rightarrow$ P}) that
these objects obey, at least in a statistical sense (Constantin \&
Vogeley 2006; Schawinski et al. 2007; Constantin et al. 2008).  This
sequence traces trends in (1) increasing host halo mass, (2)
increasing environmental density, (3) increasing central BH mass and
host stellar mass, (4) decreasing BH accretion rate, (5) aging of the
stellar population associated with their nuclei, and (6) decreasing in
the amount of dust obscuration, which might translate into a decrease
in the amount of material available for star-forming or accretion.
This sequence therefore suggests a process of transformation of
galaxies from SF via AGN to quiescence, which may be the first
empirical evidence for an analogous duty cycle to that of the high $z$
bright systems (i.e., quasars).  State of the art hydrodynamical
models provide clear support for such a scenario, by showing that
during mergers, the BH accretion peaks considerably {\it after} the
merger started, and {\it after} the star-formation rate has peaked
(e.g., Di Matteo et al. 2005, Hopkins et al. 2006).  Constraining the
nature of this {\it H II $\rightarrow$ S/T $\rightarrow$ L} sequence
at $z \approx 0$ will improve our understanding of the degree to which
the LLAGN phenomenon fits into the galactic BH accretion paradigm.

The X-ray emission is, arguably, the most sensitive test for accretion
and its intensity and efficiency, and thus, it is of great interest to
test and validate this sequence against large homogeneous X-ray
selected samples.  The Chandra Multiwavelength Project (ChaMP; Green
et al. 2004) provides a unique opportunity for this, by providing the
largest-area Chandra survey to date, which, when cross-matched with
the SDSS, provides an unprecedented number of galaxies in the local
Universe for which we can combine and contrast measurements of the
X-ray and optical emission.  The sample of $\sim 110$ Chandra X-ray
detected nearby galaxies (excluding broad line objects) analysed
in this study represents a significant improvement in both sample size
and homogeneity both for X-ray selection and optical spectral type
coverage.


Previous studies of the relation between the X-ray nuclear emission,
optical emission line activity and black hole masses provide important
physical constraints to the LLAGN phenomenon.  Almost invariably, the
conclusions are that LLAGN are probably scaled down versions of more
luminous AGN \citep{ho01, pan06}, and that $M_{\rm BH}$ is not the
main driver of the (soft) X-ray properties \citep{gree07}.  The LLAGN
are claimed to be X-ray detected at relatively high rates, and are
found to be relatively unabsorbed, obscuration appearing to play only
a minor role in their detection rates and/or in classifying them as
types 1 and 2 in X-rays \citep{rob00, hal01, min08, ho08}, with the
exception of those known to be Compton thick.    Nonetheless, the X-ray
investigations of AGN activity at its lowest levels remain largely
restricted to Ls and Ss.

Deciphering the ambiguous nature of Ls in particular, has been the
target of many X-ray studies of LLAGN focused on these sources 
\citep{yaq95, ish96, iyo98,iyo01,ter98,ter00,
  ter00b,ter02,pell00a,pell00b,pell02,georg02,ptak96, ptak99, ptak04,
  rob01}.  A hard, power-law AGN signal is generally
$spectrally$ resolved for the majority of them, however, the corresponding
energy (photon) index is marginally steeper (softer) than in
(broad line) Ss; many of them require a soft thermal component, that
somehow differs from the blackbody soft excess commonly seen in Ss and
quasars.  The Fe K$\alpha$ emission or the Compton reflection component
are usually weak in these sources, indicating that X-ray reprocessing
is not by material in an optically thick accretion disk
\citep{light88, geo91}.  Because most of these studies are based on
large-beam observations, mostly $ASCA$ or $BeppoSAX$, Ls' emission has
also quite often been attributed to stellar processes.  Higher spatial
resolution $Chandra$ and $XMM-Newton$ observations \citep{bohr01,
  kim03, pell03, ter03, filho04, page04, star05, flo06, gon06, sor06}
remain torn between these findings, as the stellar interpretation 
persists for quite a number of these sources.

In this work we approach the LLAGN phenomenon via the 
{\it H ~II $\rightarrow$ S/T $\rightarrow$ L $\rightarrow$ Passive}
  galaxy evolutionary sequence described above.  In particular, we
test the validity of the sequence within X-ray selected LLAGN via
a large variety of optical emission properties, which first provided
evidence for the sequence, and also in their X-ray properties.  We
combine the ChaMP X-ray detections with a sample of SDSS DR4 nearby
galaxies that excludes broad line objects, creating a large sample
of galaxy nuclei that spans a
range of spectral types, from passive to actively line emitting
systems, including the star-forming and accreting types, along with
those of mixed or ambiguous ionization.  Through measurements of their
X-ray spectra\footnote{i.e., the shape of the spectral energy distribution
quantified by the photon index $\Gamma$ that best fits a power-law
$N(E) \propto E^{-\Gamma}$}, fluxes and luminosities, we characterize
the sequence in terms of strength and mode of accretion.  We provide
here the first investigation of the relation between $\Gamma$ and the
Eddington ratio $L/L_{\rm edd}$ at the lowest levels of accretion.  We
reveal a rather surprising anti-correlation between these two measures,
opposite to what luminous AGN
and quasars exhibit.  This finding reveals a turning point in the
general $\Gamma - L/L_{\rm edd}$ relation followed by AGN, which is
identical (within the errors) to that shown by the $\Gamma - L/L_{\rm
  edd}$ trends in black hole X-ray Binaries (XRBs).

Throughout this work we assume $\Omega_m = 0.3$, $\Omega_{\Lambda} =
0.7$, and $H_0 = 70h$ km s$^{-1}$ Mpc$^{-1}$.

\section{The ChaMP-based LLAGN Sample} \label{sample}

The sample of LLAGN employed in this study is obtained by
cross-matching the SDSS DR4 spectroscopic sample of galaxies with the
X-ray detected sources identified as part of the $Chandra$
Multiwavelength Project (ChaMP).  ChaMP is a wide-area serendipitous
X-ray survey based on archival X-ray images of the $|b|> 20$ deg sky,
obtained with Chandra's AXAF CCD Imaging spectrometer (ACIS).  A
summary of the survey is presented in ~\citet{gre04} and
~\citet{gre09}, while ChaMP results and data can be found at {\tt
  http://hea-www.harvard.edu/CHAMP}.  The X-ray analysis extends to a
total of 392 fields through $Chandra$ Cycle 6, that cover a total of
$\sim 30$ deg$^2$ of sky area.

We limit the investigation to the DR4 dataset in order to employ the
measurements of absorption and emission line fluxes and equivalent
widths (EW) drawn from the catalog built by the MPA/JHU
collaboration\footnote{publicly available at {\tt
    http://www.mpa-garching.mpg.de/SDSS/} \citep{bri04}}.  Here the
line emission component is separated and subtracted from the total
galaxy spectrum based on fits of stellar population synthesis
templates ~\citep{tre04}.  The catalog does not include broad-line
objects.  To relate the central BH accretion activity to the host
properties, we employ stellar masses of galaxies and the H$\delta_A$
Balmer absorption-line index as a proxy for the age of the associated
stellar population, as calculated and presented by \citet{kau03}.  A
detailed analysis of these properties, and their relation to the AGN
phenomenon revealed through optical signatures, are presented in
\citet{kau04}.

The cross-match of all ChaMP sky regions imaged by ACIS with the SDSS
DR4 spectroscopic footprint results in a parent sample of 15955
galaxies on or near a chip, and a subset of 199 sources that are X-ray
detected.  Among those, only 107 sources have an off-axis angle (OAA)
$\theta < 0.2$ deg. and avoid $ccd=8$ due to high serial readout
noise; these objects comprise the main sample we employ for this
study.
Subsequent subsections present details of the X-ray spectral analysis,
together with a presentation of their general optical and X-ray
properties, the definition of the subsamples based on their
optical spectral properties, and a discussion of the selection effects
associated with these samples.


\subsection{X-ray  spectral analysis} \label{xanalysis}

$Chandra$ imaging with ACIS provides energy resolution sufficient to
constrain the X-ray spectral properties as well.  To characterize the
X-ray activity of the ChaMP-SDSS galaxies included in our sample, we
perform direct spectral fits to the counts distribution using the full
instrument calibration, known redshift and Galactic 21cm
column\footnote{Neutral Galactic column density $N^{Gal}_H$ taken from
  \citet{Dickey90} for the $Chandra$ aimpoint position on the sky.}
$N^{Gal}_H$.  Source spectra are extracted from circular regions with
radii corresponding to energy encircled fractions of $\sim 90\%$,
while the background region encompasses a 20\arcsec\, annulus,
centered on the source, with separation 4\arcsec, from the source
region.  Any nearby sources are excised, both from the source and the
background regions.

The spectral fitting is done via {\tt
  yaxx}\footnote{http://cxc.harvard.edu/contrib/yaxx} 
\citep{Aldcroft06}, an automated script that employs the CIAO
$Sherpa$\footnote{http://cxc.harvard.edu/sherpa} tool.  Each spectrum
is fitted in the range 0.5 -- 8 keV by two different models: (1) a
single power-law plus absorption fixed at the Galactic 21cm value:
(model "{\tt PL}"), and (2) a fixed power-law of photon index
$\Gamma = 1.9$ plus intrinsic absorption of column $N_H$ (model "{\tt
  PLfix}").  These fits use the Powell optimization method, and
provide a robust and reliable one-parameter characterization of the
spectral shape for any number of counts.  Spectra with less than 100
net counts were fit using the ungrouped data with Cash statistics
\citep{Cash79}, while those with more than 100 counts were grouped to
a minimum of 16 counts per bin and fit using the $\chi^2$ statistic
with variance computed from the data.  For the (9) objects with more
than 200 counts we employ a third model where both the slope of the
power-law and the intrinsic absorption are free to vary (model "{\tt
  PL\_abs}").  

Many of the X-ray detected galaxies in our sample have relatively few
net counts (mean 76, median 19).  In such cases, instrumental hardness
ratios is often used, in the belief that genuine spectra fitting is
not warranted by the data quality.  We stress however that spectral
fitting provides the most consistent and robust estimates of the
physical parameters of interest, the power law slope and intrinsic
absorption.  Because the ChaMP X-ray exposures span a variety of
intervening Galactic columns, include data from both back- and
front-side ACIS CCDs, and span 6 years of observations, the spectral
response between sources varies significantly.  While constraints from
spectral fitting may not be tight for low count sources, use of
unbinned event data and the appropriate response gives an optimal and
unbiased estimate of the fit parameters and their uncertainties,
especially important when absorption may be present at different
redshifts.  Classical hardness ratio analysis on the other hand
amounts to grouping the data into two rather arbitrary bins,
introducing potential biases and statistical complexity.  Interpreting
the hardness ratio value for ChaMP sources in disparate fields
requires incorporating the instrument response in any case, so we
strongly prefer spectral fitting.  Nevertheless, when there is only one free parameter,
only the overall spectral shape is constrained.

This simple parametrization proves generally sufficient to model the
0.5 -- 8 keV spectra of these objects.  Comparisons of 0.5 -- 8 keV
fluxes $f_x$ (and luminosities $L_X$) obtained from the {\tt PL} and
{\tt PLfix} models show good agreement, for the whole sample of galaxies,
the average (median) of the difference in these values being
0.07 (0.01) dex.  We caution that the simple power-law fits we use
here could be misleading for objects where the absorption is complex
(i.e., a partial covering, with one or more absorber potentially being
ionized).  However, the data quality is insufficient  to show that the situation 
is more complex than a simple power-law.


We compile a set of "best" measurements for $\Gamma$ or $N_H$, by
using the values obtained from the {\tt PL} (intrinsic $N_H$ fixed at zero) and
{\tt PLfix} ($\Gamma$ fixed at 1.9) models, respectively.  For objects with more than 200
counts we use the $\Gamma$ and $N_H$ values obtained from the {\tt
  PL\_abs}.  The mean $\Gamma$ for the whole sample of 107 galaxies is
2.03 $\pm$1.38, with a median of 2.04.  The level of intrinsic
absorption is generally low.  More than 85\% of the sample exhibits
$N_H < 1 \times 10^{22}$ cm$^{-2}$, while for 60\% of the objects the
spectral fits are consistent with zero intrinsic absorption.  Note that, given the simplified model 
used in fitting the X-ray spectra, these values might not necessarily represent the true distribution 
of absorption in these objects. 
Individual measurements of all of these X-ray properties, together
with their observational parameters, like the total number of X-ray
counts, the exposure time, the off-axis angle, together with their
corresponding X-ray source ID, are listed in Table ~\ref{table-xray}
for all 107 objects.


Contribution from thermal emission is expected for some of
the objects included in this sample of nearby galaxies, whether or
not they show line emission activity.  Such a component may provide a
reasonable contribution to the total (X-ray) emission even in objects
where the dominant ionization mechanism, as identified optically, is a
compact nuclear source, i.e., an AGN. 
LINERs, for instance, have frequently been associated with
photoionization by hot, young stars \citep{fil92, shi92, bar00},
clusters of planetary nebula nuclei \citep{tan00}, or more recently
(and perhaps more consistent with their older stellar populations), by
hot post-AGB stars and white dwarfs \citep{Stasinka08}.  

We attempted fitting the $0.5 - 8$ keV
spectra with a Raymond-Smith (R-S) thermal plasma model, with the
abundance fixed at 0.5 solar.  The choice of abundance level is
inspired by previous investigations of LLAGN, e.g., \citet{ptak99} and
\citet{ter02}, even though in many
cases the abundance remains poorly constrained.  This R-S model
fit results seem physically feasible for only about the third of the sample.
Reasonable values, in agreement with previous findings, i.e.,
$kT \la 2$ keV are only found for 35 sources.  For another third the
best fit $kT > 10$ keV.  We discuss in more detail the results of
fitting this model as a function of the optical spectroscopic
properties of these sources in Section~\ref{xgen}.

As probably expected, it is the passive galaxies that show 
the softest power-law slopes we measure ($\Gamma>3.5$),
suggesting that, for these cases in particular, a power-law
representation may be incorrect.  If we instead fit a Raymond-Smith model, we
derive reasonable typical temperatures near $\sim 0.7$ keV.
Nonetheless, since
even these objects are likely to have some power-law contribution
from X-ray binaries, the R-S model is perhaps no better a
characterization of the true spectrum than a power-law.
The X-ray flux distribution we derive from the R-S model fits to
passive galaxies is not significantly different ($\pm 25\%$),
so we prefer to retain the power-law fits everywhere to facilitate
more direct comparison of the different spectral classes.

\subsection{The Optical  Spectral Classification} \label{oclass}

We identify and classify accretion sources and other types of active
systems in both the parent galaxy sample and the X-ray detected
subsample, based on their optical emission line properties.  It has
been argued \citep{ho97s, con06, kew06} that the best way to separate
accretion sources from starbursts or other types of active systems is
via a set of three diagnostic diagrams, which employ four line flux
ratios: [\ion{O}{3}]$\lambda$5007/H$\beta$,
[\ion{N}{2}]$\lambda$6583/H$\alpha$,
[\ion{S}{2}]$\lambda\lambda$6716,6731/H$\alpha$, and
[\ion{O}{1}]$\lambda$6300/H$\alpha$.  Thus, for both samples, we first
select a subset of strong emission-line sources that show significant
emission in all six lines used in the type classification (H$\alpha$,
H$\beta$, [\ion{O}{3}],[\ion{N}{2}], [\ion{S}{2}], and [\ion{O}{1}]),
and a set of passive objects that show insignificant line emission
activity.  An emission feature is considered to be significant if its
line flux is positive and is measured with at least 2$\sigma$
confidence.  Following ~\citet{kew06} classification criteria, the
emission-line objects are separated into Seyferts (Ss), LINERs (Ls),
Transition objects (Ts), and star-forming, or H {\sc ii} galaxy
nuclei.

This method of classifying low luminosity actively line-emitting
galaxy nuclei has the disadvantage that it leaves unclassified a high
fraction ($\sim$ 40\%) of galaxies, which show strong emission
features, but not in all six lines considered here.  The condition for
strong emission in [\ion{O}{1}] in particular is significantly
restrictive.  Moreover, another quite large ($\sim$ 25\%) fraction of
the emission-line objects remains unclassified, as their line ratios,
although accurately measured, do not correspond to a clear spectral
type in all three diagrams.  In the majority of such cases, while the
[\ion{N}{2}]/H$\alpha$ ratio shows relatively high, S-like values, the
corresponding [\ion{S}{2}]/H$\alpha$ and/or [\ion{O}{1}]/H$\alpha$
place them in the T or H {\sc ii} -like object regime; thus, because
the [\ion{S}{2}] and [\ion{O}{1}] emission lines are better
AGN-diagnostics than [\ion{N}{2}], these systems are likely to be
excluded from the AGN samples selected via these classifications.  As
a consequence, our samples based on the 6-line classification are
small.

\begin{figure}
\plotone{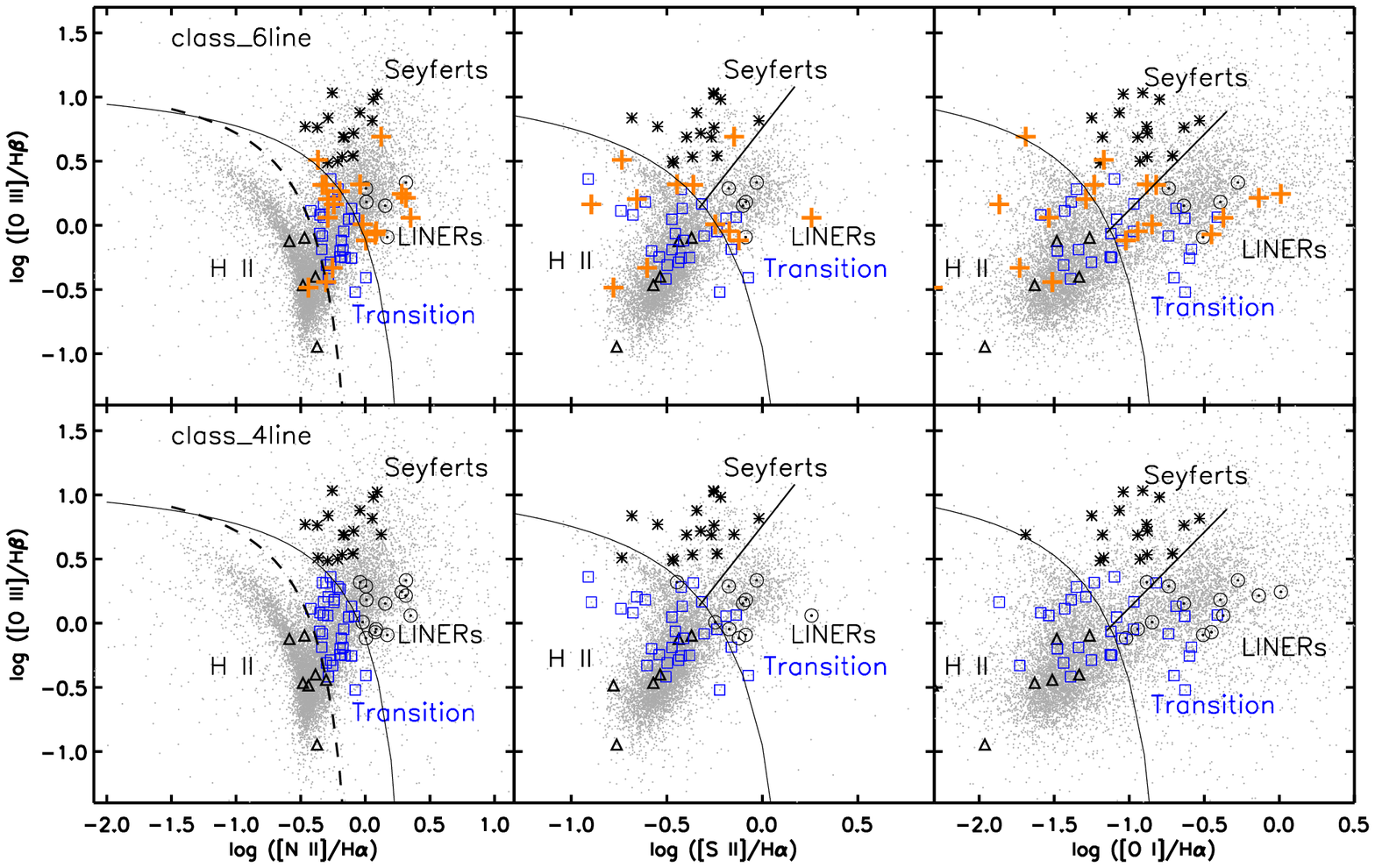}
\caption{Diagnostic diagrams for all ChaMP galaxies that exhibit
  emission-line activity, with relatively high ($> 2$) signal-to-noise
  line flux measurements in all 6 lines ({\it top}), and in only
  [\ion{N}{2}], [\ion{O}{3}], H$\beta$, and H $\alpha$
  ({\it bottom}).  The solid and dashed black 
 curves illustrate the ~\citet{kew01} and ~\citet{kau03} separation
 lines while the diagonal
  lines illustrate the separation between Seyferts and LINERs by
  ~\citet{kew06}.  The grey points correspond to galaxies in ChaMP
  fields that fall on or near the ChaMP ACIS chips, but are not
  necessarily X-ray detected.  Seyferts, Liners, Transition Objects,
  and H {\sc ii} galaxies are shown as asterixes, open circles, blue
  squares, and triangles, respectively.  The orange crosses are
  6-line non-classified objects with a 4-line class.  It is quite
  apparent that the combination of the 4-line classification and X-ray
  detection is very efficient in distinguishing between different
  types of emission among LLAGN, quantitatively identical to the
  6-line method.
  \label{bpt}}
\end{figure}

To enlarge our samples of galaxy nuclei of all spectral
types, we also explored an emission-line classification based on only
the [\ion{O}{3}]/H$\beta$ vs.  [\ion{N}{2}]/H$\alpha$ diagram, i.e., a
4-line classification method, for the X-ray detected sources.  The
emission line galaxy samples comprise thus all objects showing at
least 2$\sigma$ confidence in the line flux measurements of these four
lines only.  The delimitation criteria of H {\sc ii}'s and T's remain
unchanged, while Ss and Ls are defined to be all objects situated
above the ~\citet{kew06} separation line, and with
[\ion{O}{3}]/H$\beta$ $>3$ and $<3$ respectively.  The 4-line and
6-line classifications result in significantly different classes when
applied to optically selected galaxies in general \citep{con06}.  In
particular, true properties become heavily blended into the dominant
population of LINERs (or Ts, depending on the separation lines used in
the diagnostic diagrams).
Interestingly, however, when applied to
the X-ray sample, the 4-line classes fall well within the 6-line
loci;  although Ss and Ls are separated only by their
[\ion{O}{3}]/H$\beta$ line flux ratio, they remain clearly separated into
the [\ion{S}{2}]/H$\alpha$ and [\ion{O}{1}]/H$\alpha$ diagrams as well.
Figure ~\ref{bpt} shows how the 6-line
(top) and the 4-line (bottom) classifications compare for the ChaMP X-ray
detected galaxies.

Although the sample of X-ray detected galaxies is small, this
comparison indicates that adding X-ray detection makes the 4-line
classification more secure, and that the need for the (usually unavailable)
[\ion{O}{3}]/H$\beta$ vs. [\ion{S}{2}]/H$\alpha$, and
[\ion{O}{3}]/H$\beta$ vs. [\ion{O}{1}]/H$\alpha$ diagrams is not as
stringent as in the cases where only optical information is available.
We will thus consider for the analysis presented in this
paper only the 4-line classification.

\subsection{The X-ray detection fraction of the LLAGN} \label{statistics}

The "cleaning" role of the X-ray detection in finding and defining
LLAGN is even more obvious when we compare the fractions of X-ray
detected objects by spectral type, both relative to the parent
sample of nearby optically selected objects and the subsample of X-ray
detected galaxies.  Table ~\ref{stats} lists these 
percentages, where for the parent optically selected sample we
consider only the SDSS galaxies on ACIS chips (excluding chip S4,
$ccd \# 8$), and with off-axis angle $\theta < 0.2$\,deg, consistent
with the conditions used in compiling the X-ray samples.  The first
two columns show the number (and fraction) that each spectral type
represents, among (1) the optical parent sample and (2) the
X-ray-detected subsample. The third column lists the raw fraction of
X-ray detections per spectral type. 

That the X-ray detection is very efficient in finding LLAGN,
particularly Seyferts, is quite apparent.  While narrow-line Ss
usually make up only $<2 - 4\%$ \citep{ho97d, con06} of the optically
selected nearby galaxies, the X-ray detection increases the chances of
finding them tenfold.  Ts and Ls, where an AGN contribution to the
total ionization power is expected, are also much better represented
in the X-ray detected sample of galaxies, their fractions being $\sim
3 \times$ larger than when only optical selection is employed.  Some
65\% of the optically defined Ss are detected in X-rays, while the
other spectral types hardly reach an X-ray detection fraction of
$20\%$.  Ls are the second most X-ray active sources within nearby
galaxies, while Ts and the galaxies that show some/weak emission line
activity account for less than 1/5th of the sample.  As we discuss
further in Section ~\ref{xgen}, only the luminous H{\sc ii}s are
detected in X-rays.  While the X-ray detection rate of the H {\sc ii}s
is basically consistent with zero, when detected, their X-ray emission
is moderately strong, $L_X = 10^{39} - 2.5 \times 10^{41}$ erg
s$^{-1}$; half of the H {\sc ii} detections show X-ray luminosities
higher than the level that can be reached without contribution from
AGN ($10^{40}$ erg s$^{-1}$), suggesting once more that the nuclear
emission in these sources might not be completely driven by stellar
processes.

The only previous studies that encompass the whole spectral variety of
LLAGN are \citet{rob00} and \citet{par08}, which employ ROSAT 
(HRI and RASS respectively).  Their search for X-ray
emitting nearby galaxy nuclei, optically characterized via the Palomar
and SDSS surveys respectively, concluded in soft X-ray detection rates
of $\sim 70\%$ of both Ss and Ls (HRI), $\sim 70\%$ of Ss and $\sim
60\%$ Ls (RASS), $\sim 40\%$ (HRI) and $< 10\%$ (RASS) of H {\sc
  ii}'s, and $\sim 30\%$ of passive galaxies (both cases).  Comparison
of these detection rates to ours is problematic, because of their
softer instrument bandpass, lower sensitivity, but wider sky
coverage.  

Hard X-ray studies of homogeneously selected
samples including all spectral types of nearby active galaxies are
practically non-existent.  Ls, and particularly those found in the
Palomar survey, have been clearly privileged in terms of X-ray
targeting \citep{ho01}.  Their detection rates are found to be
significantly higher than what we report here based on the
serendipitous ChaMP survey.   Ho et al. (2001) reports a $\sim 70\%$
detection rate, while, when chosen for having a flat-spectrum radio
core, Ls are found to be $100\%$ X-ray active \citep{ter03}.  Later
studies claiming better accounting for selection and classification as
LINER \citep{sat04, dud04, pel05, sat05, flo06, gon06} conclude with 
lower fractions, $\sim 50\%$.  

ChaMP has the distinct advantage of presenting a large, homogeneous
serendipitous sample of LLAGN.  The detection fraction is not an
intrinsic property of galaxies, but rather a convolution of galaxy
properties with optical survey depth, and the X-ray sensitivity
vs. sky area curve.  As described in \citep{gre09}, the ChaMP is
characterising the X-ray sensitivity at the position of every SDSS
galaxy, which will enable us to compile the unbiased fraction of
galaxies by optical spectral type (e.g., Ls) that fall in X-ray
luminosity bins, from sky volumes complete to those limits.  We will
present the results of such an investigation in a subsequent paper.

\subsection{Selection Effects: the ChaMP X-ray Galaxy Sample is
  Minimally Biased} \label{selection}

X-ray and optical emission are correlated.  Thus, while the X-ray
selection is one of the most powerful tools to exploit in detecting
accretion sources, it is also expected that this selection picks up,
selectively, the brightest (optical) sources.  Due to its
serendipitous character, ChaMP should however reduce such effects.
Note that, out of 107 X-ray detections that this ChaMP sample of galaxies
provides, only 13 are targets.

\begin{figure}
\plottwo{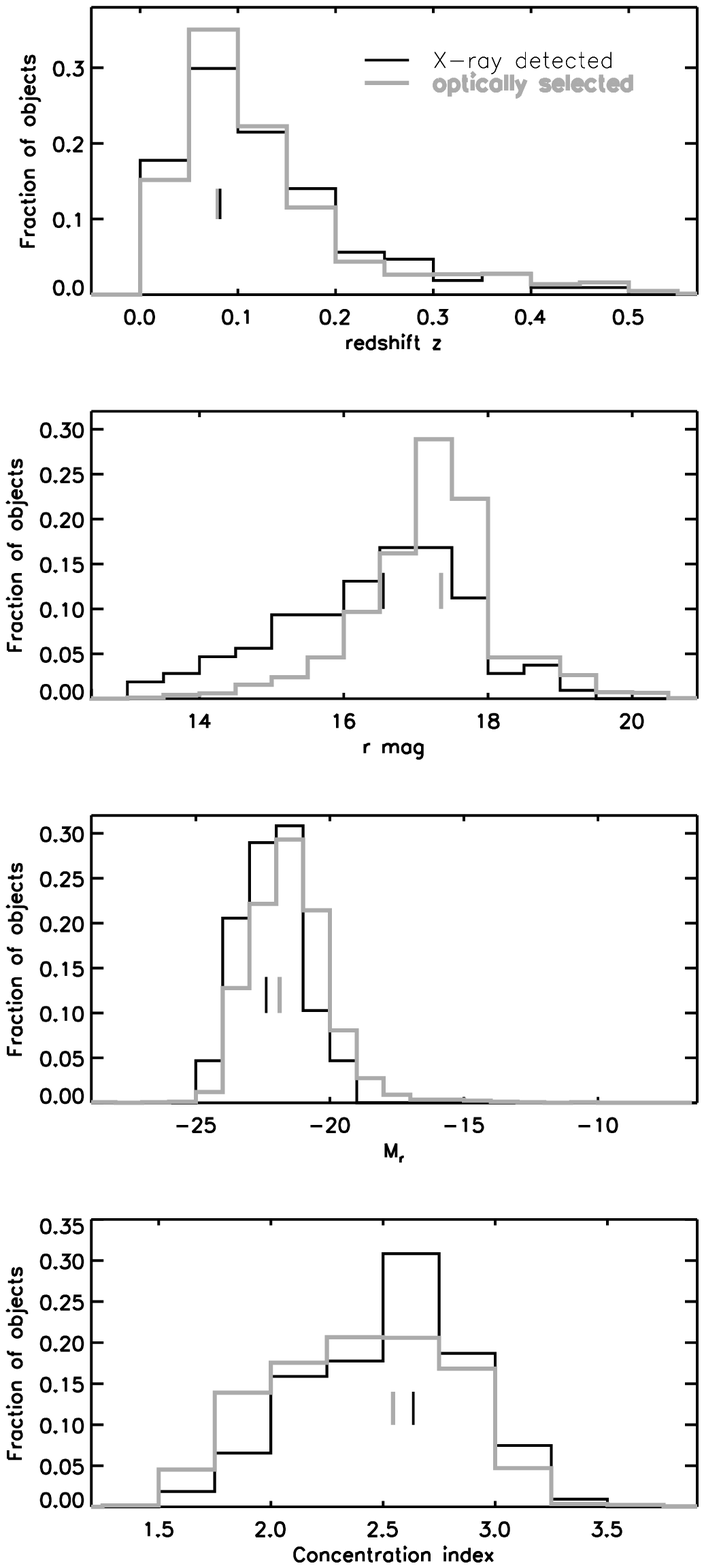}{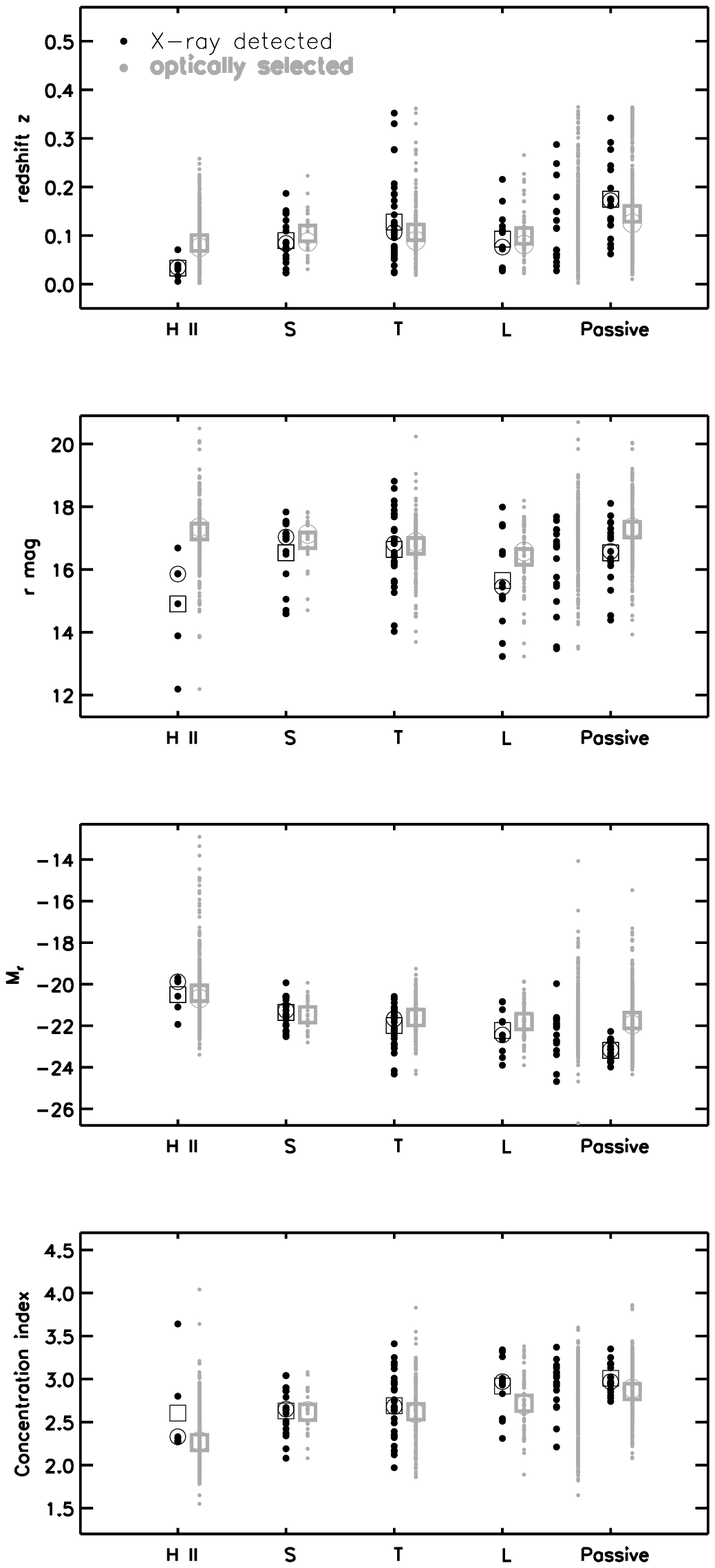}
\caption{Comparison of the optical properties of X-ray detected
  galaxies and all optically selected (SDSS) galaxies on or near ChaMP
  fields/chips. {\it Left:} 
  Histograms of redshift ($z$), apparent $r$-band magnitude ($m_r$),
  absolute $r$-band magnitude ($M_r$), and the concentration index
  ($C$) for the full
  samples of SDSS galaxies on/near ChaMP fields and the ChaMP
  detections; median values are indicated by the vertical bars.  {\it
    Right:} Individual measurements of $z$, $m_r$, $M_r$ and $C$ are
  shown separately per galaxy spectral type; both average (squares)
  and median (circles) values are indicated here, for all types of
  objects.  The bias caused by the X-ray  detection criterion is
  clearly weak; the few X-ray detected H{\sc ii}'s dominate the
  observed difference in $r$ and $M_r$ distributions. 
  \label{host}}
\end{figure}

We explore in Figure ~\ref{host} the biases that X-ray
detection potentially adds to our sample.  Comparisons of the
distributions of redshift $z$, apparent and absolute $r$-band (SDSS)
magnitudes, $r$ and $M_r$ respectively, and of the concentration
index $C$\footnote{$C$ = $R_{50}/R_{90}$, where $R_{50}$ and $R_{90}$
  are the radii from the center of a galaxy containing 50\% and 90\%
  of the Petrosian flux} as a proxy for the morphological type of
these galaxies.  Both for the whole optically and X-ray selected
samples (histograms on the left panel) and separately per spectral
type (the right panel) show, pleasingly, that biases are not strong.

However, H {\sc ii} galaxies are of lower $z$ and brighter $M_r$ when
detected in X-ray.  As shown in the next section, the H {\sc ii}s are
generally weak X-ray sources.  The tendency for X-ray detected
galaxies to appear brighter in apparent magnitude ($r$, by $\sim 0.5$
mag) seems to be caused by the large difference in $r$ between the
H{\sc ii} galaxies alone, as all the other types of sources show very
similar ranges, averages or medians, when analysed separately.

The concentration index $C$ appears to be somewhat larger for the
X-ray objects.  Several factors are likely to account for this.  The H
{\sc ii}s are the least concentrated optically, and have the lowest
X-ray detection fraction.  X-ray detection is sensitive to the AGN
activity, which is more prevalent in the early type (massive
bulge-dominated) galaxies \citep{ho03, kau04}.  Nuclear activity is
also expected to increase $C$ simply by adding light to the core.  Note
that there is no significant difference in this parameter in regard to
Ss' X-ray selection.  However, Ss make up for a tiny fraction of $z
\approx 0$ galaxies, and the morphology of their hosts spans quite a
range.

\subsection{General X-ray Properties in Relation to Optical
  Spectroscopic Classification} \label{xgen}

\begin{figure}
\plottwo{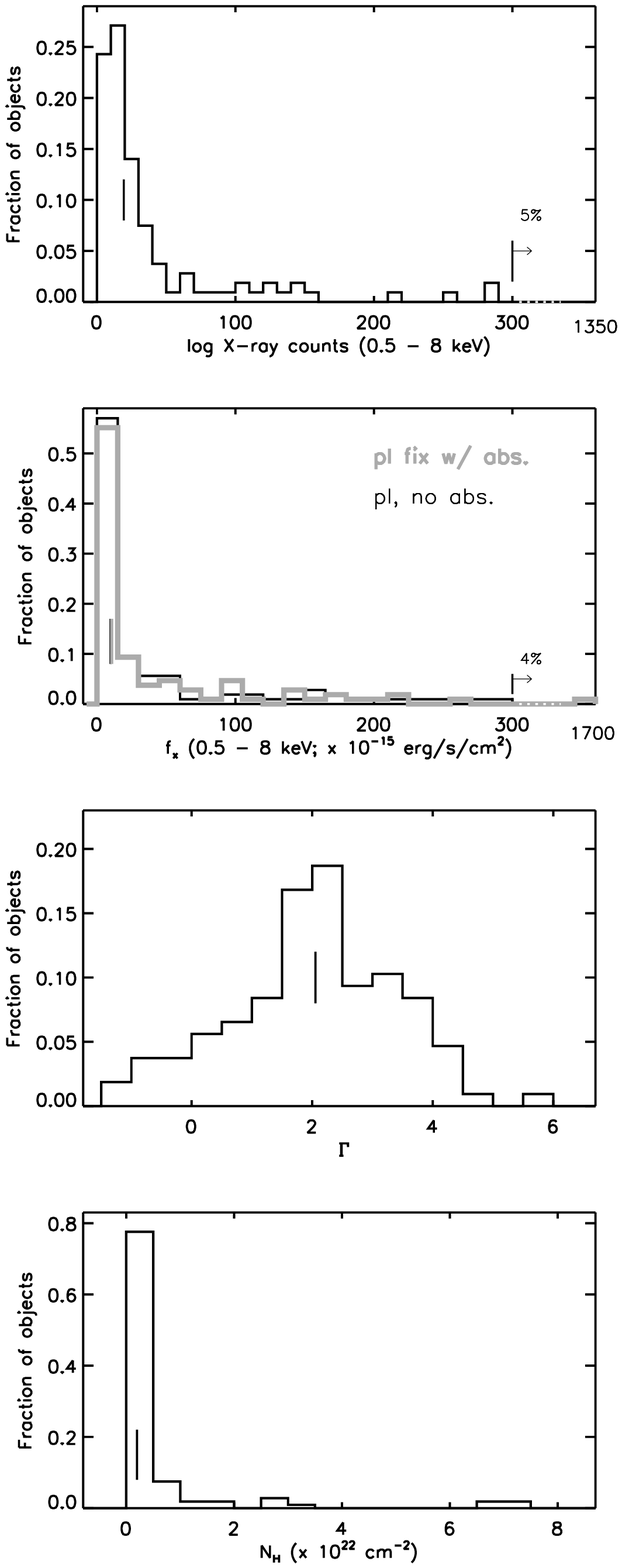}{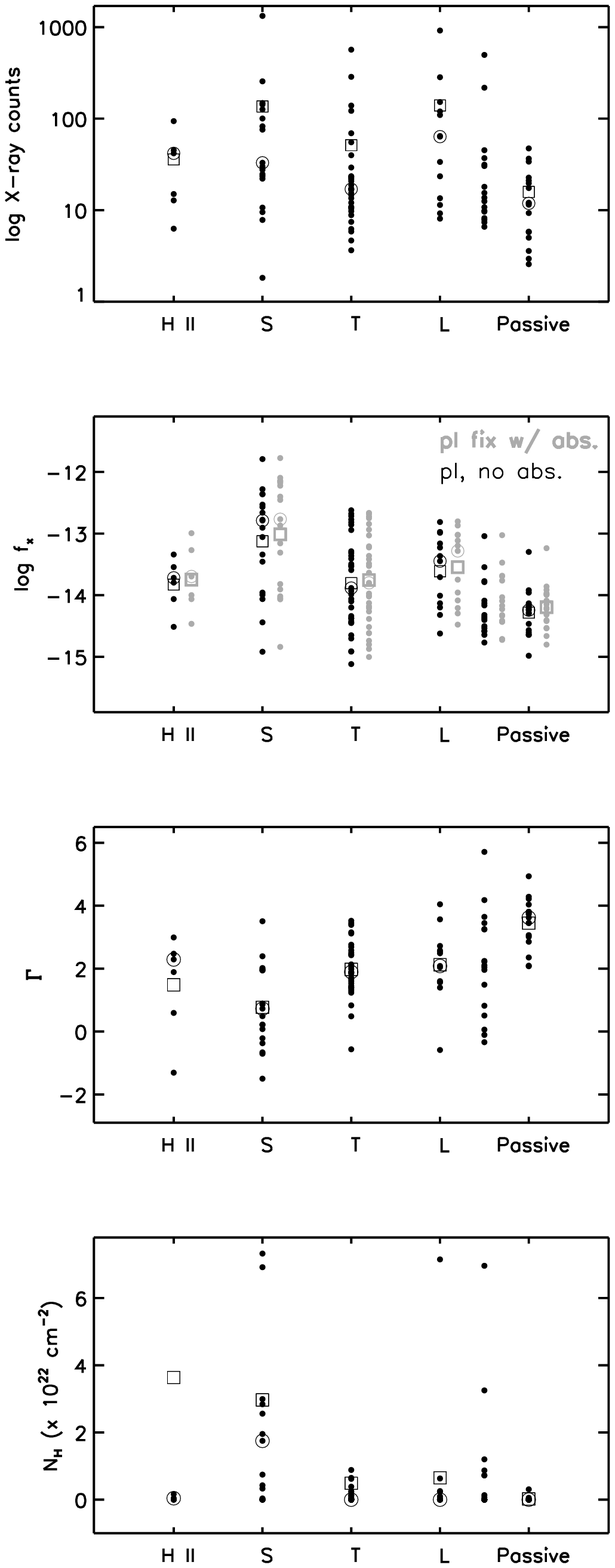}
\caption{Distribution of X-ray properties of the ChaMP galaxies. {\it
    Left:} Histograms of X-ray counts in the 0.5 -- 8 keV regime, the
  X-ray flux $f_x$ (calculated via two fitting models, a power-law
  with no intrinsic absorption and a fixed power-law with variable
  absorption), the (best) X-ray  photon index $\Gamma$, and the (best)
  $N_H$ (see text); median values are indicated by the vertical bars.
{\it Right:} Individual measurements of X-ray counts, $f_x$, $\Gamma$,
  and $N_H$ shown separately per galaxy spectral type; 
  average and median values are indicated for all types of objects by
  squares and circles respectively.   
  \label{hostx}}
\end{figure}

X-ray information about the nearby galaxy centers constrains the
contribution of accretion-related processes to their optical spectral
characteristics.  We present in this section the distributions of a
variety of X-ray properties for the whole sample as well as for
subsamples by optical spectral class.  Figure ~\ref{hostx} shows the
distributions of the X-ray counts, the $0.5 - 8$\,keV unabsorbed X-ray
fluxes, the best-fit X-ray photon indices $\Gamma$ and the intrinsic
neutral hydrogen column densities $N_H$.  These
measurements are shown for the whole sample of X-ray detected nearby
galaxy nuclei (left column), and separately per spectral type (right
column).

For X-ray fluxes, we show the values derived using the primary
power-law fitting models discussed in Section ~\ref{xanalysis}: (1) a
single power-law with no intrinsic absorption and (2) a fixed
power-law ($\Gamma=1.9$) with absorption.  The $f_x$ values are
generally consistent with each other for the whole range of brightness
and optical spectral type.  As expected, the X-ray brightest objects
are among Ss, however, even for this spectral type the range of values
remains pretty broad, spanning 3 orders of magnitude.  Note also that
the few H {\sc ii} galaxies that are X-ray detected are in general
brighter than the passive systems.

In terms of the X-ray spectral shape, the ChaMP nearby galaxies are
quite a diverse population.  The mean photon index per optical
spectral type shows however a rather clear dependence on the spectral
type: Ss show the hardest $0.5 - 8$\,keV spectral shape, becoming
softer and softer from Ts to Ls to the Passive galaxies which are
clearly the softest.  The H {\sc ii} galaxies are unexpectedly hard in
average $\Gamma = 1.46$, however, two particularly hard detections
clearly weight the subsample in this direction.  The Ts and Ls average
at $\Gamma \approx 2$.

The $N_H$ values are poorly constrained for this sample, and there is
no obvious correlation with the optical spectral type.  It is however
obvious that the Ss are the sources with the highest fraction of
non-zero absorption.  Since all our Ss are of type 2, i.e., lack broad
emission lines in their spectra, the unification paradigm predicts
that many will show signature of absorption in X-rays.  A typical
unabsorbed power-law $\Gamma = 1.9$ requires a column density of $N_H
= 8 \times 10^{21}$ cm$^{-2}$ to yield an apparent $\Gamma = 1$
similar to the mean value found for our S subsample, which is 
consistent with the observed mean $N_H$ for these particular systems.

\section{Probing The Sequence} \label{sequence}

The {\it H {\sc ii} $\rightarrow$ $S$/$T$ $\rightarrow$ L}
evolutionary sequence proposes a comprehensive picture for the
co-evolution of AGN and their host galaxies.  This scenario is
supported by and strengthens previous studies of AGN, star-formation
activity and their co-evolution in nearby galaxies \citep{kau04,
  hec04}, and may enhance our understanding of how AGN work and evolve
in relation to both their hosts and their environments \citep{con08}.

X-rays, as primary signatures of supermassive BH accretion, offer a
critical verification of the proposed sequence.  $L_X$ and spectral
fits characterize the sequence in terms of strength and mode of
accretion, especially the order of $S$s and $T$s within the {\it H
  {\sc ii} $\rightarrow$ $S$/$T$ $\rightarrow$ L} cycle.  Both the
bulge nebular properties, and the small and large scale environments
of $S$s and $T$s, are very similar and remain intermediate between
those of H {\sc ii}s and $L$s.  The only parameters showing a ``jump''
in the otherwise smooth trends are H$\alpha$/H$\beta$ Balmer
decrements and the nearest neighbor distance $d_{\rm 1nn}$
\citep{con08}.  H$\alpha$/H$\beta$ provides a measure of absorption,
and perhaps also the amount of fuel available for accretion, which we
can now test directly against both $L_X$ (accretion power) and X-ray
spectral constraints.

\begin{figure}
\plotone{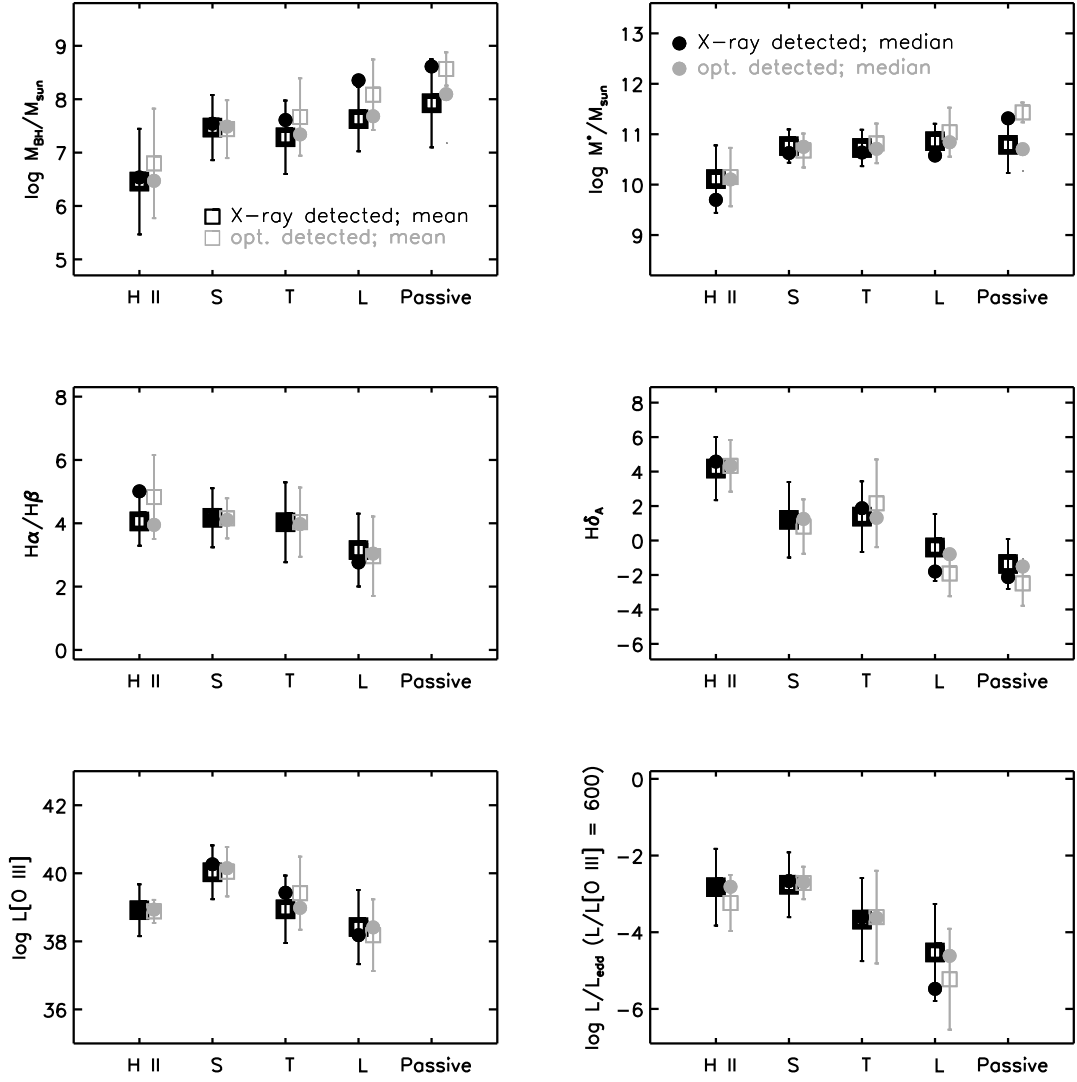}
\caption{Comparison of median and average optical properties of X-ray detected galaxies
  and all optically selected galaxies on or near ChaMP
  fields/chips,  along the {\it H {\sc ii} $\rightarrow$ S
    $\rightarrow$ T $\rightarrow$ L $\rightarrow$ passive galaxies}
  sequence. The plots show: the BH mass (based on $M_{\rm BH} -
  \sigma_*$ relation of Tremaine et al. 2002), the (dust corrected)
  stellar mass (log $M_*/M_{\sun}$), the Balmer decrement, the
  H$\delta_A$ Balmer absorption-line index, $L[{\rm O III}]$, and the
  accretion rate expressed by $L_{\rm bol}/L_{\rm Edd}$, where $L_{\rm
    bol}/L[O III] = 600$; average and median values are indicated by
  squares and circles respectively.   
  \label{seqo}}
\end{figure}

We present in Figure ~\ref{seqo} a comparison of the degree to which
optically and X-ray selected galaxies follow the proposed sequence in
terms of the black hole mass $M_{\rm BH}$, obtained via $\sigma_*$
measurements and the $M_{\rm BH} -\sigma_*$ relation \citep{tre02}, the
(dust corrected) stellar mass $M^*$ \citep{kau03a}, the Balmer
decrement H$\alpha$/H$\beta$ as a proxy for dust extinction, the
H$\delta_A$ Balmer absorption index as a measure of the age of the
associated stellar population, $L[O III]$, and the accretion rate
expressed as $L/L_{\rm edd}$, where $L = L_{\rm bol} = 600 \times L[O
III]$ for the bolometric correction \citep{hec04, kauffmann09}, and
the $L[O III]$ is extinction-corrected using the corresponding Balmer
decrements and a $\tau \propto \lambda^{-0.7}$ attenuation law
\citep{charlot00}.  Given the (spectral) definition of passive
galaxies (i.e., lacking optical emission line activity), there are no
measurements of the Balmer decrement, $L[O III]$ and $L/L_{\rm edd}$
for these sources.  It is notable that in all measures, and for
all types of sources, the X-ray and optically selected sources are
very similar, and appear to obey the {\it S $\rightarrow$ T
  $\rightarrow$ L $\rightarrow$ P} sequence.

If anything, the sequence appears stronger among the X-ray selected
galaxies, both in median/average values and in their distributions of
individual measurements, which span smaller ranges of values than for
the optically selected objects.  Among the weakest accreting objects,
Ls and Passive galaxies, the X-ray selection tends to pick up more
massive objects, with heavier BHs, and older stellar populations;
there is however no obvious difference in these parameters for the
other types of sources.  As expected, these massive systems also
appear to have smaller [\ion{O}{3}] luminosities and accretion rates,
accentuating the sequential $S$ to $T$ to $L$ drop in these parameters
suggested by optically selected/defined objects.  Also, while the
Balmer decrement distributions suggest that, in general, the X-ray
selection is not strongly affected by dust, the slightly more obscured
X-ray detected H {\sc II}s and less obscured X-ray detected Ls make
the sequence more apparent.

\begin{figure}
\plotone{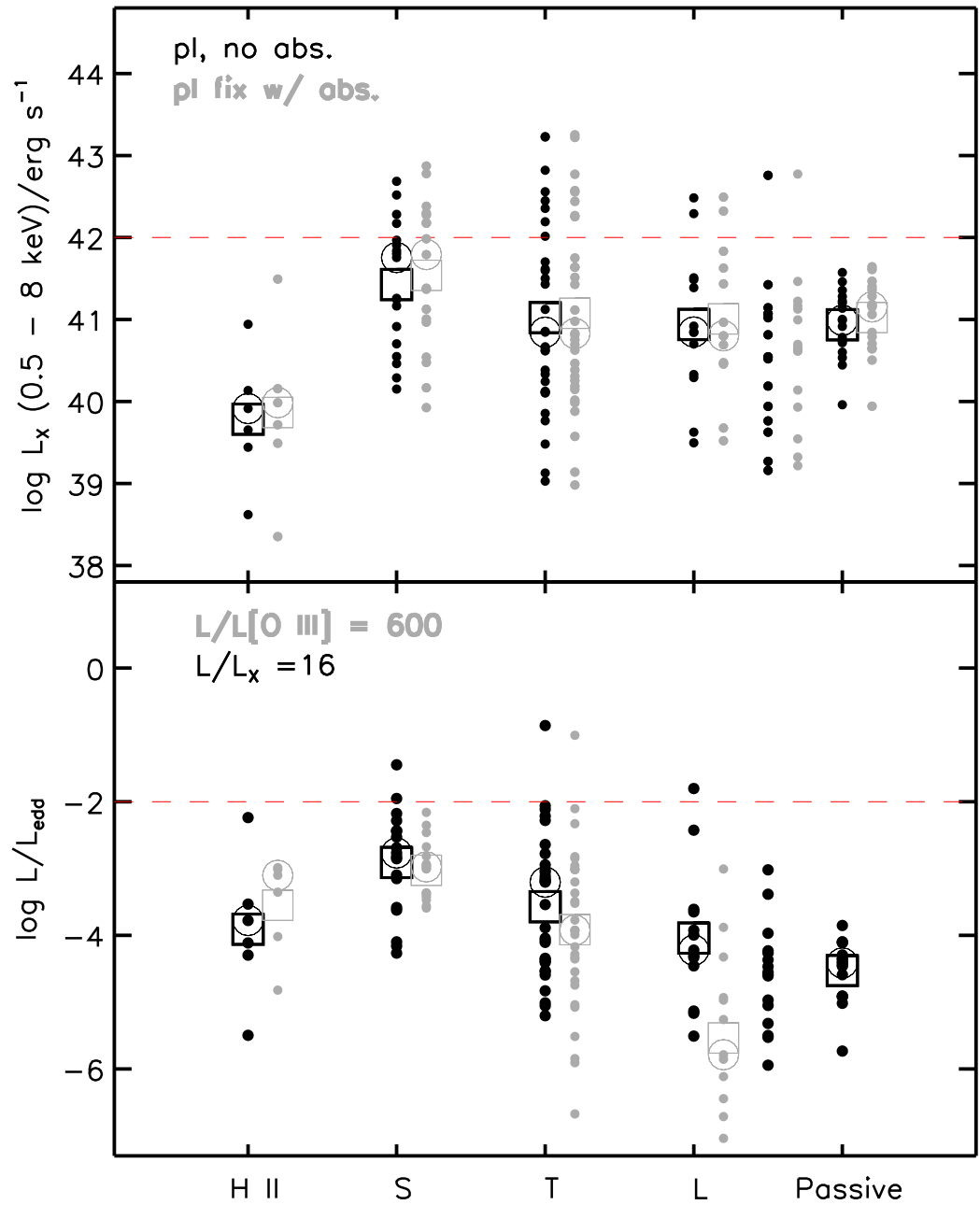}
\caption{The {\it H {\sc
      ii} $\rightarrow$ S $\rightarrow$ T $\rightarrow$ L
    $\rightarrow$ passive galaxies} sequence in X-ray: the X-ray
  luminosities obtained via $yaxx$ for two fitting models (1. a
  power-law with no intrinsic absorption and 2. a fixed power-law with
  $\Gamma=1.9$ with intrinsic absorption); the accretion rate
  values, expressed by $L/L_{\rm Edd}$, calculated using both $L_{\rm
    bol}/L_X = 16$ and $L_{\rm bol}/L[O III] = 600$ are compared;
  average and median values are indicated by squares and circles 
  respectively.  
\label{seqx}}
\end{figure}

Figure ~\ref{seqx} illustrates how the 0.5 - 8 keV X-ray luminosity
$L_X$ and the corresponding $L/L_{\rm edd}$ behavior along the
{\it S $\rightarrow$ T $\rightarrow$ L $\rightarrow$ P} sequence.  For
the sake of comparison, we show $L_X$ values obtained via two spectral
fitting models, 1. a power-law with no intrinsic absorption (only the
Galactic one) and 2. a fixed power-law with $\Gamma = 1.9$ and with
variable intrinsic absorption added to the Galactic level.  The
sequence is supported by both types of measurements, albeit stronger
when the simple power-law model is used.  As discussed in Sections
~\ref{xanalysis} and ~\ref{xgen}, the free power-law with no intrinsic
absorption spectral model, that seems to best characterize this sample
of objects, makes the statistical sequence even more probable.  For
the sequence in the accretion rate, we calculate $L/L_{\rm edd}$ using
$L/L_X = 16$ for the bolometric luminosity, as suggested by
\citet{ho08}, and the BH masses estimated from their host stellar
velocity dispersion $\sigma_*$ using \citet{tre02}.  We also contrast
the X-ray measurements with those where $L_{\rm bol}$ derived from $L
[O III]$; note that while $L_X$ and the derived $L/L_{\rm edd}$ are
available for the whole sample of 107 objects, only 69 of them exhibit
strong $2-\sigma$ detectable, [\ion{O}{3}] line emission.  The $S
\rightarrow T \rightarrow L \rightarrow P$ galaxy sequence compares
well in both optical and X-ray measurements.  The values of all these
parameters decrease monotonically in both median and average values,
from Ss to Ts, to Ls and Passive galaxies, consistent with what
optical properties of these sources put forward.

The H {\sc ii}s are the only apparent exception here.  The small
number statistics for these galaxy nuclei preclude any strong
conclusions.  Given the expected high sensitivity to soft sources that
these measurements provide, their spectra tend to be on the hard side.
Obscuration is a notorious cause for spectral hardening, and thus, the
possibility that these objects hide in their centers obscured BH
accretion is still not ruled out.

Note also that the power-law slope $\Gamma$ that best fits the X-ray
spectra increases from Ss to Ts, to Ls, in both median and average
values, showing a tendency of softening of the spectra from Ss to
passive galaxies along the proposed sequence (Figure ~\ref{hostx}).
This is quite an interesting finding, as in other AGN, mostly the
luminous type 1 AGN, observations suggest an opposite trend: the
stronger accreting (and more luminous) sources are the softer ones,
most recently quantified by \citet{kelly08, she08}.  It is interesting
to note that a spectral softening with strengthening of the accretion
process/rate is also a generally common feature of the X-ray Binaries
(XRBs) with reasonably high $L/L_{\rm edd}$ \citep{kub04}.  Note
however that, when the Eddington ratio is less than a critical value,
$L/L_{\rm edd} \la 0.01$, i.e., XRBs are observed in their low/hard
state, there is a clear trend for softening with further weakening of
the accretion rate \citep{kalemci05, yamaoka05, yuan07}.  We
investigate this finding in more detail in the following section, and
discuss the analogy with the XRB phenomenon.

\section{Dependence of $\Gamma$ on $L/L_{\rm edd}$} \label{gamma}

Investigations of how X-ray parameters depend on the accretion rate
relative to the Eddington rate are expected to offer important
constraints on physical models of the AGN X-ray emitting
plasma, particularly its geometry.  A hot, optically thin
corona that Compton up-scatters UV photons from the optically thick
disk \citep{sha73, haa91} seems to fit reasonably well the X-ray
emission of the highly accreting systems, particularly when it is
associated with a hot, possibly patchy and "skin"-like structure
"sandwich"-ing the cold disk \citep{nay00, cze03}.  Other geometries
remain however viable, among them an accretion disk evaporating into a
hot inner flow \citep{sha76, zdz99}, or combinations of a hot inner
flow and the patchy corona \citep{pou97, sob04}.  For LLAGN, with
$L/L_{\rm edd} < 10^{-9} - 10^{-5}$, the accretion flow has been
hypothesized to originate from a geometrically thick and hot disk-like
structure that is inefficient at converting gravitational potential
energy into radiation, the radiatively inefficient accretion flow
(RIAF) model \citep{nar94, bla99, nar00}.  Models suggest that there
is a transition/switch from a standard disk to an advection dominated
accretion flow (ADAF; or, a radiatively inefficient accretion flow,
RIAF) when $L/L_{\rm edd}$ declines below a critical value within a
certain transition radius \citep{esin97,yua04,lu04}.  In either case,
radiation pressure driven outflows can also alter the physics of the
corona \citep{pro05}.  Because the efficiency in producing an X-ray
accretion flow and in driving the outflows depends on the BH mass and
its accretion rate, it is important to understand the inter-dependence
of these parameters on the X-ray properties, particularly the shape of
the X-ray spectrum, i.e., the X-ray photon index $\Gamma$.

The relationship between $\Gamma$ and the Eddington ratio $L/L_{\rm
  edd}$ is relatively well studied, and yet a controversial issue.
These two parameters seem to be positively correlated for objects
accreting at relatively high Eddington ratios, i.e., quasars, luminous
type 1 Seyferts \citep{kelly08, she08}, while for low $L/L_{\rm edd}$
values the situation remains uncertain, mainly due to the lack of
quality data in that regime.  
The conclusion so far seems to be that the shape of the hard
X-ray power-law is largely controlled by $L/L_{\rm edd}$.  For the
luminous, strongly/efficiently accreting sources, it is proposed that
the corona acts as a "thermostat" by (Compton) cooling more
efficiently when the disk emission increases, producing more soft
photons, and thus steepening the hard X-ray spectrum.  This scenario
also accounts nicely for the generally narrow range of $L/L_{\rm edd}$
and $\Gamma$ values measured in (optically selected) quasars or
luminous AGN, in general [$L/L_{\rm edd} \sim 0.3$ with a typical
dispersion of a factor of $\sim 5$; \citep{mcl04, kol06, net07,
  shen08}.  $\Gamma \sim 1.5 - 2.5$; \citep{vig05, she08}].  For
LLAGN, the $\Gamma - L/L_{\rm edd}$ relation remains only vaguely
constrained.

\subsection{$\Gamma - L/L_{\rm edd}$ relation for nearby
  AGN/galaxies} \label{llagncorrel}

\begin{figure}
\plotone{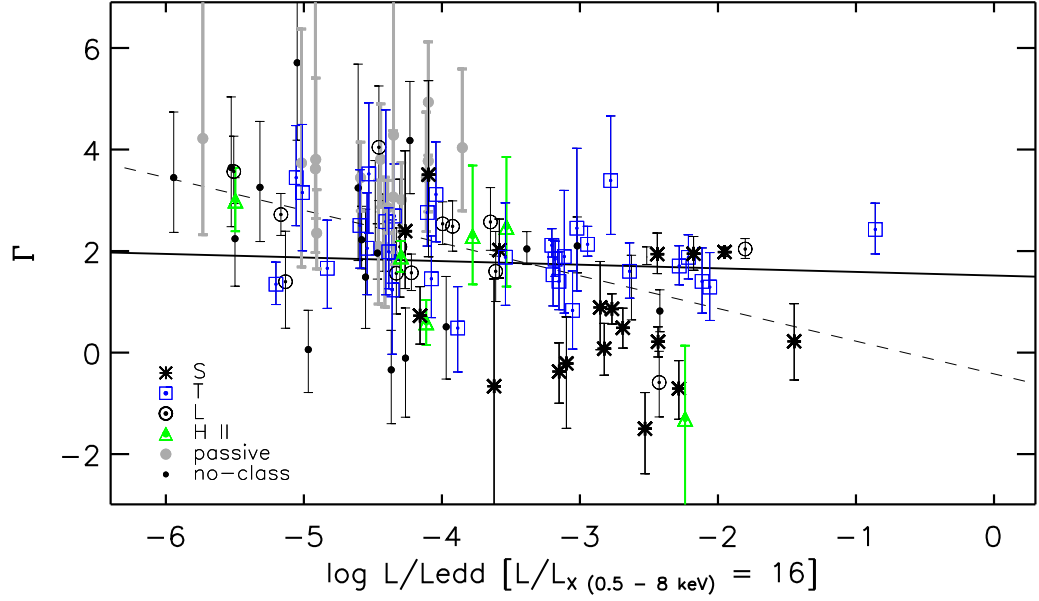}
\caption{Distribution of the X-ray photon index as a function of
  $L/L_{\rm edd}$ where $L = L_{\rm bol}$ is estimated using
  $L_{\rm bol}/L_X = 16$.  There is clearly apparent an
  anti-correlation between these two measures; the symbols reflect the
  different spectral optical classification.  The solid and dotted
  lines reflect the best fit linear relations with the errors weighted
  and not weighted, respectively.  
  \label{gamma_edd}}
\end{figure}

The relation between the X-ray photon index $\Gamma$ and the Eddington
ratio for the ChaMP X-ray detected galaxies is illustrated in Figure
~\ref{gamma_edd}.  The Eddington luminosity $L_{\rm edd}$ is
calculated as indicated in Section ~\ref{sequence}, using $M_{\rm BH}$
values estimated based on stellar velocity dispersion $\sigma_*$ via
\citet{tre02}, while the bolometric luminosity is calculated based on
$L_X$, using the average bolometric correction of $L_{\rm bol}/L_X =
16$ \citep{ho08}.  There is a rather clear trend of spectral hardening
with increasing accretion rate.  $\Gamma$ and $L/L_{\rm edd}$ are
found to be negatively correlated.

Defining an accretion rate, i.e., calculating $L/L_{\rm edd}$, for
galaxies with $L_X < 10^{42}$ erg s$^{-1}$ may be misleading if the
X-ray emission in these objects is dominated by X-ray sources other
than an accreting super massive BH (e.g., individual compact binaries or hot
diffuse gas).  Note however that the $x$-axis of
Figure~\ref{gamma_edd} is simply the measure of $L_X/M_{\rm BH}$ (or
better, $L_X/{\sigma_*}^4$), where both $L_X$ and the BH mass $M_{\rm BH}$
(or $\sigma_*$) are measured in exactly the same manner for all of the
objects involved, and the trend remains even if this parameter is not
interpreted as an accretion rate.  The strongest likely dilution to
accretion emission comes from contributions of the hot ISM in passive
galaxies.  Hot gas emission is soft, and relatively stronger in more
massive hosts, both of which would push the passive galaxy points
toward softer X-ray spectra (larger $\Gamma)$ and lower $L_X/M_{\rm
  bh}$, which might spuriously accentuate the observed trend even in
the absence of significant accretion power.  Indeed, the observed
trend is weakened once the passive galaxies are removed
(Table~\ref{lsf}).  So while we rule out significant extended emission
contributions in this sample, more detailed examination of such
objects is warranted to determine the relative fractions of nuclear
vs. extended hot gas contributions.

We measure the significance of the $\Gamma - L/L_{\rm edd}$
anti-correlation using the Spearman-rank test.  The Spearman-rank
coefficient, the chance probability, and the number of sources for
each correlation are listed in Table ~\ref{spearman}.  We fit the
anti-correlation points with a linear least-squares method for the
whole sample and for the subsamples of galaxies corresponding to
different spectral types, using the {\tt
  mpfit}\footnote{http://www.physics.wisc.edu/\~craigm/idl/fitting.html}
routine, being able to account for the errors in $\Gamma$.  We show,
for comparison, both the error weighted (continuous line) and the
unweighted (dotted line) best-fits in Figure ~\ref{gamma_edd}.  The
scatter around the best linear fit is large, and as expected the
points with the largest error bars show the largest deviation.
However, within the errors, the results of the fit remain unchanged
when we use only objects with small measurements errors (i.e., $\Delta
\Gamma/\Gamma < 50\%$).  The results for the linear regression
coefficients and the corresponding $\chi^2$ and $dof$ values are
listed in Table ~\ref{lsf}.  For the sake of considering ``cleaner''
AGN-like activity only, we also list here the results of such a
fitting techniques to samples that exclude the Passive galaxies and
the H {\sc ii}s, and for samples of luminous X-ray systems ($L_X \ga
10^{42}$ erg s$^{-1}$) only, and indications of an anticorrelation,
albeit weaker, remain.

The linear regression fits might appear at odds with the conclusion of
the Spearman test, which indicates that $\Gamma$ and $L/L_{\rm edd}$
are possibly anticorrelated for all the subsamples presented here.
Note however that such a discrepancy appears for the subsamples where
the Spearman test remains rather inconclusive as, the probability that
an anticorrelation appears by chance is large.  Moreover, the Spearman
test ignores the errors, and thus tests the unweighted data, for which
linear regression fits are always consistent with negative slopes.  It
is clear however that for Seyferts in particular, there is no evidence
for either positive or negative correlation between $\Gamma$ and
$L/L_{\rm edd}$.  This has been seen in other samples as well
\citep{winter09}, and it is an important result.  Nevertheless,
investigations of the LLAGN phenomenon should not be restricted to
these types of sources only.

\begin{figure}
\plotone{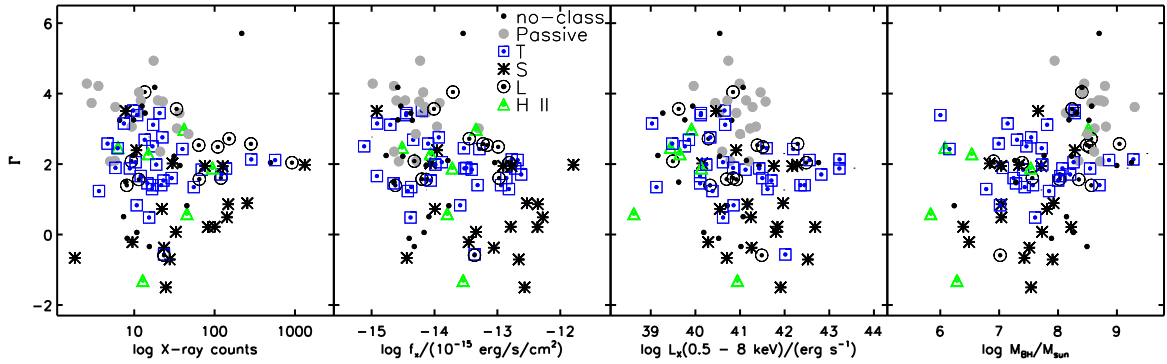}
\caption{The relation between the $0.5 - 8$ keV X-ray photon index and
  the total number of counts, the $0.5 - 8$ keV flux $f_x$, the $0.5 - 8$ keV
  luminosity $L_X$, and the BH mass.  No
  relation appears to exist between $\Gamma$ and X-ray counts, while
  there are rather weak apparent negative correlations with
  $f_x$ and $L_X$; $M_{\rm BH}$ does not appear to correlate
  with $\Gamma$.
  \label{gammal}}
\end{figure}

In an attempt to provide some more physical insights into the reality
and significance of this new trend, we also explore here the possible
correlations between various X-ray measures that might influence (if
not artificially create) it.  In particular, we scrutinize the way our
measured $\Gamma$ relates to the number of counts, the X-ray flux and
luminosity.  Some studies showed that the photon index correlates with
the X-ray luminosity, becoming softer in more luminous sources,
whether measured in soft, $0.2 - 2$ keV \citep{forster96, lu99,
  gierlinski04, williams04}, or hard, $2-10$ keV \citep{dai04,
  porquet04, wang04} bands only, or comprising the whole spectrum, and
even as they vary \citep{chiang00, petrucci00, vaughan01}.  However,
because many others have not found such trends, the idea that the
choice of the sample involved in these studies may contribute to
producing the correlations has also been put forward.  Interestingly,
our ChaMP sample does not show this correlation either.  Figure
~\ref{gammal} illustrates the dependence of $\Gamma$ measured in the
$0.5 - 8$ keV range as a function of the total number of counts in
this energy range, the X-ray flux, and the X-ray luminosity.  As
before, we emphasize the different optical spectral classifications;
the fact that, e.g., the various types of galaxies separate rather
well from each other in diagrams like this suggests that the
correlation is physical, and most probably related to the 
accretion physics in these nuclei, rather than an artificial effect
of fitting various (simple) models to their X-ray spectra.

\subsection{Comparison with high $z$ QSOs} \label{qsocorrel}

\begin{figure}
\plotone{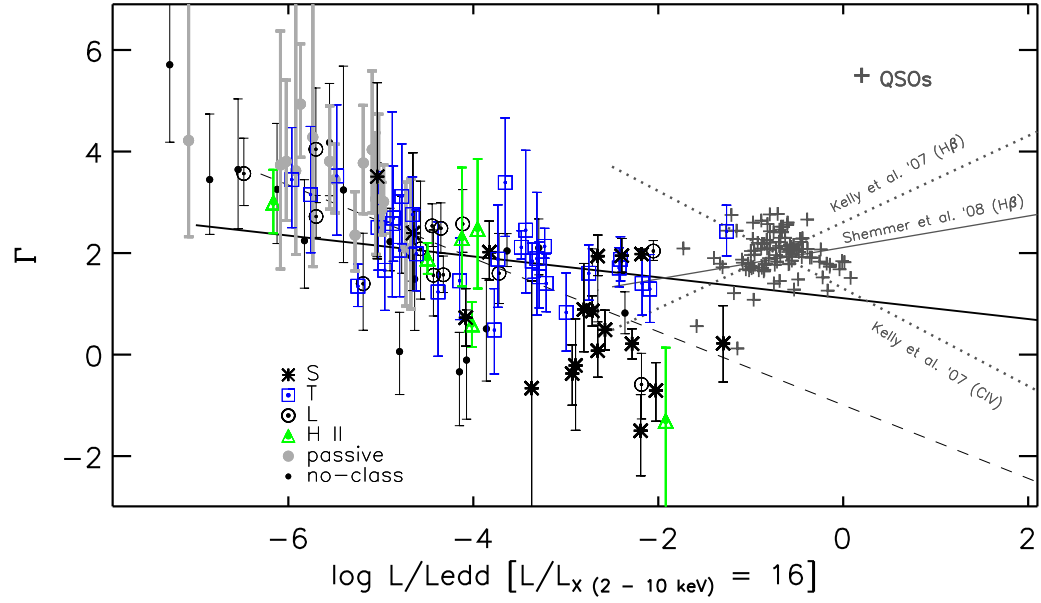}
\caption{Distribution of the X-ray photon index $\Gamma$ as a function of
  $L/L_{\rm edd}$ where $L = L_{\rm bol}$ is estimated using
  $L_{\rm bol}/L_X = 16$, with $L_X=L_X(2-10 keV)$ calculated from
  $L_X$(0.5 - 8 keV) using $\Gamma$.  We show for comparison the
  measurements of $\Gamma$ and $L/L_{\rm 
    edd}$ for the ChaMP (high $z$) SDSS QSOs with spectra
  \citep{gre09}, where both $\Gamma$ and $L/L_{\rm edd}$ are
  calculated in the same way as for the ChaMP galaxies/LLAGN, with the
  difference that the bolometric correction for quasars is considered
  to be $L_{\rm bol}/L_X = 83$, independend of their luminosity.
  \label{gamma_edd_qso}}
\end{figure}

The anti-correlation between $\Gamma$ and $L/L_{\rm edd}$ (or simply,
$L_X/M_{\rm BH}$) that we find to characterize the nuclear emission of
nearby galaxies is certainly surprising.  This trend is opposite to
what more luminous galaxy nuclei, i.e., quasars, exhibit: their X-ray
spectra soften as they become more luminous.  Figure
~\ref{gamma_edd_qso} shows the $\Gamma - L/L_{\rm edd}$
anti-correlation followed by the low luminosity galaxy nuclei along
with measurements of $\Gamma$ and $L/L_{\rm edd}$ for a sample of SDSS
quasars with optical spectra that are ChaMP detected and
X-ray analysed in the same manner we handled our sample of nearby
galaxy nuclei \citep{gre09}.  The quasar $L/L_{\rm edd}$ values are
obtained using the average bolometric correction of $L/L_X = 83$
estimated by \citet{ho08}, with no luminosity-dependence, and the
black hole masses from \citet{shen08}.

To ease comparison with previous work on quasars, we use in this plot
the $2 - 10$ keV $L_X$ luminosity, which is obtained by extrapolating
the measured unobscured $L_{\rm 0.5 - 8 keV}$ value, using the best
$\Gamma$ measurements.  Note that the anti-correlation between
$\Gamma$ and $L/L_{\rm edd}$ that nearby galaxy nuclei show is even
more pronounced when the $2 - 10$ keV $L_X$ is plotted against
$\Gamma$.  This is expected, as softer objects will be less luminous
in $2-10$ keV than in the $0.5 - 8$ keV range, while the hard objects
will be more luminous.  The
magnitude of this effect, i.e., the ratio of the two
$L_X$ values is a function of the photon index $\Gamma$, as
given by

\begin{equation}
L_X(2 - 10\ keV) = L_X(0.5 - 8\ keV) \times \frac{10^{2-\Gamma} - 2^{2-\Gamma}}{8^{2-\Gamma} - 0.5^{2-\Gamma}}.
\end{equation}

The largest difference, and thus the most significant effect on the
shape of the $\Gamma - L/L_{\rm edd}$ trend, is for the softest
($\Gamma \ga 3$) objects, however it does not exceed $\sim1$ dex.
Note that for these particular objects the measurement errors are also
among the highest, and hence contributed the least weight to the
best-fit $\Gamma - L/L_{\rm edd}$ relation.

The ChaMP quasars fall well within the locus of values expected based
on previous work.  The ChaMP quasars do not show, however, a clear
$\Gamma - L/L_{\rm edd}$ positive correlation.  The ChaMP quasars
span a wide range in redshift and their $L/L_{\rm edd}$ measurements
reflect a mix of BH mass estimates based on all H$\beta$, \ion{Mg}{2},
and \ion{C}{4}, which may add scatter to an underlying correlation
\citep{shen08}.  We show for comparison the results of linear
regression fits of the $\Gamma - L/L_{\rm edd}$ correlation reported
by \citet{she08} and by \citet{kelly08}, with the results for the
$H\beta$ and \ion{C}{4} based estimates of the $M_{bh}$ shown
separately.  Note that we use here these data only for the purpose of
global comparison of the quasar properties with those of the nearby
galaxy nuclei, and do not attempt to improve upon the previous work on
the characterization or calibration of the $\Gamma - L/L_{\rm edd}$
relation for quasars.

Clearly, the quasar $\Gamma$'s are not negatively correlated with
their Eddington ratios, even with the shift in $L_X$ produced by the
energy conversion mentioned above, which might contribute to such
trend (the quasar X-ray data is analysed and measured via the same
techniques employed for the nearby galaxy nuclei).  Thus, over the
whole $L/L_{\rm edd}$ range we explore here by putting together
luminous and weak AGN, there is clearly a break in the $\Gamma -
L/L_{\rm edd}$ correlation, which seems to happen at $L/L_{\rm edd}
\approx 10^{-2}$, and $\Gamma \ga 1.5$.  It is important to note that,
for the samples of quasars investigated by both \citet{kelly08} and
\citet{she08}, the scatter in the claimed correlation between $\Gamma$
and $L/L_{\rm edd}$ is largest as the samples reach the weakest
accretion rates (while hardly reaching the $L/L_{\rm edd} \approx
10^{-2}$ level), and the hardest values ever measured for the quasar
photon index $\Gamma \ga 1.5$; while suggestive of a break, the quasar
data alone cannot however be used to single it out.  Other studies
that include measurements of lower luminosity managed to point out
these types of deviations from the expected correlation
\citep{zhang08, gre09}, however, the data remained sparse at those
levels of accretion, leaving the effect unquantified.

We note here that the comparison of the nearby galaxies' central X-ray
emission with that of the high redshift QSOs may be inadequate since
X-rays from SF activity is not yet accounted for in the former.  We
addressed this issue by including in the modelling of the original
X-ray spectrum (Section ~\ref{xanalysis}) a $\Gamma=2$ component with
the expected $L_X(SF)$.  A power-law with $\Gamma=2$ provides the best
simple description of the mix of hot gas and high mass X-ray binaries
(HMXBs) that comprise the SF activity \citep[e.g.,][]{kim92a, kim92b,
  nandra94, ptak99, george00, colbert04, reddy04, lehmer05}.  To
estimate the potential $L_X(SF)$ in each galaxy, we use star-formation
rates (SFR) calculated and available for these objects in the MPA/JHU
catalog (Section ~\ref{sample}; Brinchmann et al. 2004), along with
the $L_X - SFR$ correlation quantified by Gilvanov, Grimm, \& Sunyaev
(2004).  There are 83 objects with total $L_X$ above the line
describing the $L_X - SFR$ correlation, for which we modelled the
remaining - presumably AGN component.  This reanalysis produces
significant deviations from the $\Gamma - L_X/L_{\rm edd}$ correlation
only for 6\% of the sample (5 out of the 83 objects involved in this
analysis).  The scatter and the error bars for both $\Gamma$ and $L_X$
are only slightly increased.  The slope of the correlation flattens
but remains consistent within the errors with our previous
measurements.

\subsection{The ``break'' in the $\Gamma - L/L_{\rm edd}$ correlation
   and comparison with XRBs} \label{break}

The X-ray photon index $\Gamma$ and the Eddington ratio
 $L/L_{\rm edd}$ show a double-sloped relation, with positive and
 negative correlations above and below $L/L_{\rm edd} \approx 0.01$
 and $\Gamma \ga 1.5$ respectively.  We see now that, a given AGN
 spectral index $\Gamma$ may correspond to 
 two different luminosity levels, with the luminosity difference
 greater for sources characterized by softer spectra (Figure
 ~\ref{gamma_edd_qso}).  This break in the $\Gamma - L/L_{\rm edd}$
 relation, when studied over a large range of accretion
 power, fits well into the theoretical ideas of a transition in the AGN
 accretion mode: a standard \citep{sha73} accretion disk/corona at
 high Eddington rates, i.e., quasar phase, and an ADAF \citep{nar94}
 at low $L/L_{\rm edd}$.  This break we find in the $\Gamma - L/L_{\rm
   edd}$ relation may provide the best empirical evidence to date for
 such a transition.

 The inflection point in the $\Gamma - L/L_{\rm edd}$ relation is
 dominated by Ts, consistent with the optical nature of these systems.
 While hypothesized to be the result of mixed AGN and SF ionization
 \citep{ho03, con09}, their AGN component could be either an S- or
 L-like.  Ss appear to be the low $L$ quasar-like, efficiently
 accreting systems, while the Ls are the ADAF objects, so T's locus at
 the break region is suggestive of a switch in the accretion mode.
 The optically-defined Ts are then transition systems that map the
 X-ray inflection as well.

Note that this idea of transition in the accretion mode, from ADAF to
standard-disk as the accretion rates increases, has been first
proposed, and later developed, mainly based on results of
investigations of smaller BH mass accretors, i.e., stellar mass
black-hole X-ray binaries, or XRBs; see \citet{narayan08} for a recent
review.  Interestingly, the relation between $\Gamma$ and $L/L_{\rm
  edd}$ that XRBs exhibit in the different phases of their temporal
variability is also multivalued: a given spectral index $\Gamma$ may
correspond to two different luminosity levels, with the luminosity
difference greater for sources characterized by softer spectra.  That
is, the XRBs show a positive $\Gamma - L/L_{\rm edd}$ correlation
while in their high/soft states \citep{kub04}, and an anti-correlation
while in their low/hard states \citep{yamaoka05}.  Several clear
examples of such a turn (or convergence point) in the $\Gamma -
L/L_{\rm edd}$ relation measured in XRBs are illustrated in, e.g.,
\citet{yuan07} and \citet{wu08}.

In terms of the physical phenomenology governing the black hole
accretion process over more than 10 orders of magnitude in the
Eddington ratio, the transition from an ADAF to a standard disk
accretion seems to adequately account for both the XRB and AGN
emission properties.  The ADAF scenario qualitatively explains the
anti-correlation via Comptonization of thermal synchrotron photons as
the dominant cooling mechanism at low $L/L_{\rm edd}$ ratios:
the Compton $y$-parameter increases with increasing of the optical
depth, which is caused by increase of the accretion rate, the X-ray
spectrum becomes harder \citep{esin97}.  Further increase in the
accretion rate would cause both an increase in the released energy and
a decrease in the electron temperature, weakening the corona, and
consequently the optical depth, reducing the $y$-parameter, and
leading to softer spectra, and thus the positive $\Gamma - L/L_{\rm
  edd}$ correlation \citep{janiuk00}.  The ADAF scenario has been
relatively successful in explaining a variety of the LLAGN properties
\citep{ho08}, while the standard disk-corona model has been widely
invoked to explain quasar emission.  Our finding of a non-monotonic
$\Gamma - L/L_{\rm edd}$ relation seems to provide the first direct
empirical link between these two different types of accretion in AGN.

The AGN-XRB physical analogy has been discussed rather extensively
\citep{mac05}, and is particularly supported by the two "fundamental
planes" of the BH activity on all mass scales, the $L_{\rm radio} -
L_X - M_{bh}$ \citep{merloni03, falcke04} and the $L_{\rm bol} -
M_{bh} - T_{\rm break}$ \citep{mchardy06}.  There was however not
much evidence for existence of spectral states in massive black holes,
similar to those of the stellar black holes.  Our discovery of an
anti-correlation between $\Gamma$ and $L/L_{\rm edd}$, and thus the
discovery of a turning point in the $\Gamma - L/L_{\rm edd}$ relation
for AGN, provides this evidence, which constitutes an important
empirical constraint to the idea that these systems are really the
analogs of each other, in spite of the vast difference of scales.

\section{Conclusions  \& Discussion}

This study of serendipitous Chandra nearby sources brings together for
the first time a large homogeneous sample of active and inactive
galaxy nuclei selected and classified based on their optical spectral
properties.  With a minimal selection bias, we characterize the X-ray
properties of low luminosity AGN via measurements of the X-ray
spectral shape, fluxes, and luminosities.  These measurements add
important information and provide new constraints to the proposed {\it
  H II $\rightarrow$ S/T $\rightarrow$ L $\rightarrow$ P} galaxy
evolutionary sequence.  Optical observations reveal that at least in
statistical terms, along this sequence, (1) the host halo mass
increases, (2) the environmental density increases, (3) both central
BH mass and stellar mass increase, while (4) the rate of accretion
onto the central Bh decreases, (5) the stellar population ages, and
(6) the material that could be used for accretion and/or
star-formation is less and less available.  The X-ray data support
these evolutionary trends and bring surprising new insights into the
nature of the LLAGN phenomenon.

There are two main results of this analysis that we want to emphasize
here: 


\begin{itemize}

\item The {\it (H II $\rightarrow$) Seyfert $\rightarrow$ Transition
    Object $\rightarrow$ LINER $\rightarrow$ Passive Galaxy} sequence
  suggested by a large variety of optical measures is supported by
  X-ray measurements.  Both the spectral shape and the accretion
  power, as measured by $L_X$ and the Eddington ratio $L/L_{\rm edd}$,
  with $L=L_{\rm bol} = 16 \times L_X$, show a clear trend toward
  softer, less X-ray luminous and less actively accreting sources from
  $S$s to $T$s, to $L$s, and, at the end, the $Passive$ galaxies.  The
  rather ambiguous (in some optical properties) succession of $S$ and
  $T$ phases is now significantly constrained by the X-ray activity to
  follow in a sense of decreasing accretion power.

\item There is a rather strong anti-correlation between the shape of
  the X-ray spectral energy distribution, quantified via the power-law
  index $\Gamma$, and $L/L_{\rm edd}$.  This finding translates into a
  break in the $\Gamma-L/L_{\rm edd}$ correlation exhibited by AGN of
  all powers, with spectral softening on either side of $L/L_{\rm edd}
  \approx 0.01$.  The transition point is
  identical to that where stellar mass BH accretors (XRBs) exhibit
  their turn in analogous $\Gamma-L/L_{\rm edd}$ trends.

\end{itemize}

The distribution of points within the observed $\Gamma-L/L_{\rm edd}$
relation exhibited by both weak and powerful AGN might have
other implications as well.  The presence of some LLAGN below the
$\Gamma \la 1.5$ turning point, which are mostly Seyferts, 
suggests that obscuration might also play a role in
shaping the observed $\Gamma-L/L_{\rm edd}$ trends at low and high
$z$.  These particular objects' rather hard photon indices must be the
indication of gas absorption of their $0.5 - 8$ keV spectra.
Their Eddington ratios huddle near the $10^{-2}$ level, being
generally weak when compared to the luminous quasars, and among the
strongest LLAGN.  Thus, the anti-correlation between $\Gamma$ and
$L/L_{\rm edd}$ shown by the nearby galaxy nuclei may well be
interpreted as a relation between absorption and accretion rate, the
objects accreting at higher rates being more obscured.  

The probably naive extrapolation of this idea at higher $L/L_{\rm
  edd}$ suggests then that more active AGN are also more absorbed.
Consequently, these increasingly absorbed systems, that would be the
type 2 ones [according to the AGN "unification" scenario that
separates (the observed appearance of) AGN in terms of orientation
relative to the line of sight] would be decreasingly likely to be
included in the (current) optically selected AGN/quasar samples.  This
interpretation is certainly consistent with the general results of
various quests for type 2 quasars: their X-ray spectra are harder than
their type 1 counterparts \citep{zakamska04, ptak06}, and they are
highly obscured \citep{zakamska05}.  The type 2 quasars are definitely
scarce compared with the type 1, and their fraction relative to that
of the type 1 ones decreases with increasing luminosity
\citep{reyes08}.  Thus, this may as well be the explanation for the
dearth of type 2 quasars.

If the type 1-2 dichotomy and consequently the "unification" are only
about the observing angle, the AGN-galaxy evolutionary sequence
suggested by the properties of the different types of nearby galactic
nuclei should be even stronger once inclination effects are removed,
as we would have a clearer view of the central engine.  On the other
hand, the 1-2 type separation, and thus the unification, might be the
result of evolution.  Some simulations suggest that for the luminous
quasars, the type 1 (unobscured) phase comes after a phase of
"blowing-out" circumnuclear matter, which might mean after the quasars
were observable as type 2.  By adding type 1 LLAGN to investigations
of the $\Gamma-L/L_{\rm edd}$ connection we might be able to
understand better the way the proposed evolutionary sequence does or
does not challenge the unification scenario.

Also, the exact location of the turning point in the $\Gamma-L/L_{\rm edd}$
relation remains to be better quantified in terms of both parameters.
Of particular caution is combining the $L/L_{\rm edd}$ measurements of
LLAGN with those of quasars, mainly because the methods used in
estimating their BH masses are not necessarily compatible.  The BH
masses of LLAGN are based on the $M_{bh}-\sigma_*$ relation, while the
(usually high $z$) quasar BH masses are obtained from the widths of
the optical broad emission lines via scaling relations; the scaling
relations are calibrated on $M_{bh}-\sigma_*$, which, recent work
suggests, does not necessarily hold at high $z$.
Another way of refining the exact location of the break lies of course
in better estimates of $\Gamma$, i.e., higher quality (higher
signal-to-noise) X-ray measurements, and/or larger samples.  The
latter alternative, in particular, seems to be feasible, as larger and
larger samples of well characterized samples of optically selected AGN
become available for cross-correlation with X-ray detections from, e.g.,
serendipitous surveys like ChaMP.

Future work (Constantin \& ChaMP 2009, in prep.) will explore the
multi-wavelength properties of a large sample of AGN that brings
together nearby LLAGN of type 2 with nearby quasars (type 1 AGN), thus
attempting to reconcile both the type 1-2 dichotomy and the problem of
mix-matching BH masses, along with providing larger statistics, and
better means of quantifying extrinsic effects such as absorption and
non-thermal processes (i.e., Comptonized emission from the accretion
disk's corona) that enable improved constraints to the analogy with
XRBs.  We will also address in this work the biases that are
potentially present in the previously claimed relationships between
the between $L_X$ and optical emission line luminosities for LLAGN,
together with the impact of using optical emission lines to estimate
$L_{\rm bol}$, and thus to the overall shape of the $\Gamma-L/L_{\rm
  edd}$ relation.

\acknowledgements 
AC thanks Christy Tremonti for valuable discussions
regarding the MPA/JHU catalog.  Support for this work was provided by
the National Aeronautics and Space Administration through {\em
  Chandra} Award Number AR7-8015A issued by the {\em Chandra} X-ray
Observatory Center, which is operated by the Smithsonian Astrophysical
Observatory for and on behalf of the National Aeronautics Space
Administration under contract NAS8-03060.

\clearpage


\LongTables

\begin{landscape}
\begin{deluxetable}{lcccccccccccccl}
\tablecolumns{15} 
\tablewidth{0pt}
\tablecaption{X-ray measurements of ChaMP-SDSS Galaxies
\label{table-xray}}
\tabletypesize{\scriptsize}
\tablewidth{0pt}
\tablehead{
\colhead{ObjID} &
\colhead{ra} &
\colhead{dec} &
\colhead{$z$} &
\colhead{srcid} &
\colhead{OAA} &
\colhead{counts} &
\colhead{Exp.} &
\colhead{Targ.} &
\colhead{$N^{Gal}_H$} &
\colhead{$N_H^{Intr.}$} &
\colhead{$\Gamma$} &
\colhead{$f_x$} &
\colhead{$L_X$} &
\colhead{Opt. Class} \\
\colhead{(1)} &
\colhead{(2)} &
\colhead{(3)} &
\colhead{(4)} &
\colhead{(5)} &
\colhead{(6)} &
\colhead{(7)} &
\colhead{(8)} &
\colhead{(9)} &
\colhead{(10)} &
\colhead{(11)} &
\colhead{(12)} &
\colhead{(13)} &
\colhead{(14)} &
\colhead{(15)}}
\startdata
 587722982300254964 & 233.296661 &   -0.756684 &  0.151 & CXOMP J153311.1-004524 &  6.8 &   27.6 &  4.9 & ... &  6.39 &  6.9$^{+ 3.6}_{-2.4}$     &  -0.7$^{+0.5}_{-0.6}$ & 218.41 & 42.94 & S \\    	  
 587722983362134277 & 206.118027 &    0.029700 &  0.135 & CXOMP J134428.3+000146 &  2.4 &    5.7 &  8.7 & ... &  1.93 & <0.8        	     	  &   2.1$^{+1.3}_{-1.1}$ &   5.14 & 41.06 & P \\   	  
 587722984431026387 & 195.126404 &    1.046143 &  0.067 & CXOMP J130030.3+010246 &  8.4 &   16.6 &  1.5 & ... &  1.63 & <1.3       	  	  &   1.3$^{+0.6}_{-0.6}$ & 152.65 & 42.05 & T \\   	  
 587725041163370794 & 174.768066 &   -1.980904 &  0.342 & CXOMP J113904.3-015851 &  3.5 &    2.5 & 14.0 & ... &  2.56 & <0.4       	  	  &   4.3$^{+5.1}_{-2.5}$ &   1.03 & 40.50 & P \\   	  
 587725470127161548 & 118.935455 &   41.204029 &  0.074 & CXOMP J075544.5+411214 & 10.1 &   33.9 &  7.2 & ... &  4.62 & <0.0       	  	  &   3.0$^{+0.7}_{-0.6}$ &  50.35 & 41.14 & P \\   	  
 587725550133117038 & 155.188263 &   63.196468 &  0.206 & CXOMP J102045.1+631147 &  4.2 &   39.8 &  6.0 & ... &  1.03 & <0.0       	  	  &   2.4$^{+0.5}_{-0.4}$ &  51.62 & 42.38 & T \\   	  
 587725550136983727 & 174.983627 &   66.098259 &  0.376 & CXOMP J113956.0+660553 &  5.6 &   15.0 &114.6 & ... &  1.17 & <0.3       	  	  &   2.8$^{+1.0}_{-0.8}$ &   0.99 & 41.19 & P \\   	  
 587725551735996593 & 127.912956 &   52.701363 &  0.058 & CXOMP J083139.1+524204 &  3.7 &   23.5 &  6.8 & ... &  3.91 & <1.4       	  	  &  -0.4$^{+0.5}_{-0.6}$ &  87.79 & 41.77 & S \\   	  
 587725591458414903 & 264.505035 &   58.503334 &  0.330 & CXOMP J173801.2+583012 &  3.3 &  138.8 &  4.5 & ... &  3.56 & <0.1       	  	  &   1.8$^{+0.4}_{-0.4}$ & 192.16 & 43.55 & T \\   	  
 587725980689301524 & 128.116867 &   52.605656 &  0.016 & CXOMP J083228.0+523620 & 11.6 &   41.7 &  9.0 & ... &  3.84 & <0.0      	  	  &   2.9$^{+0.6}_{-0.5}$ &  45.71 & 39.77 & H {\sc ii} \\  
 587726015607275687 & 183.399002 &    2.810350 &  0.132 & CXOMP J121335.7+024837 &  2.2 &   11.4 & 18.0 & ... &  1.75 & <0.8       	  	  &   1.6$^{+0.8}_{-0.8}$ &   7.37 & 41.31 & L \\   	  
 587726015607275717 & 183.441772 &    2.811480 &  0.073 & CXOMP J121346.0+024841 &  4.2 &   10.4 & 18.6 & ... &  1.75 & <0.6       	  	  &   2.0$^{+1.0}_{-0.9}$ &  13.05 & 40.90 & T \\   	  
 587726031729459313 & 220.677521 &    1.319711 &  0.033 & CXOMP J144242.6+011910 &  8.1 &  119.7 & 10.9 & ... &  3.35 & <0.2       	  	  &   1.6$^{+0.7}_{-0.5}$ & 106.12 & 41.22 & L \\   	  
 587726033305338033 & 141.248947 &    2.241570 &  0.148 & CXOMP J092459.7+021429 &  4.4 &    7.3 & 17.1 & ... &  3.73 & <3.2       	  	  &   0.5$^{+0.9}_{-1.0}$ &   8.14 & 41.53 & no-class \\ 
 587726100949238048 & 219.718658 &    3.716406 &  0.291 & CXOMP J143852.4+034259 &  9.0 &   22.6 & 54.6 & ... &  2.63 & <0.1       	  	  &   3.7$^{+1.1}_{-0.9}$ &   5.71 & 41.28 & P \\   	  
 587727179536138389 &  30.007231 &   -8.927990 &  0.052 & CXOMP J020001.7-085540 &  4.8 &   29.9 & 34.2 & ... &  2.09 & <0.2      	   	  &   2.0$^{+0.6}_{-0.5}$ &   8.64 & 40.43 & S \\   	  
 587727213347209620 & 322.206177 &   -7.787057 &  0.070 & CXOMP J212849.4-074713 & 10.1 &   12.7 & 19.5 & ... &  4.88 & 21.6$^{+51.2}_{-9.1}$     &  -1.3$^{+1.4}_{-2.5}$ &  28.58 & 41.43 & H {\sc ii} \\  
 587727223009510128 & 325.644348 &   12.505494 &  0.276 & CXOMP J214234.6+123019 &  2.6 &    5.0 & 14.0 & ... &  6.69 & <1.3        	     	  &   2.0$^{+1.3}_{-1.1}$ &   2.70 & 41.49 & P \\   	  
 587727227305394267 &  10.728508 &   -9.230717 &  0.076 & CXOMP J004254.8-091350 &  7.3 &   13.4 &  9.6 & ... &  3.53 & <0.1        	     	  &   4.0$^{+1.2}_{-1.0}$ &  19.63 & 40.24 & L \\   	  
 587727227305394310 &  10.752554 &   -9.229577 &  0.076 & CXOMP J004300.6-091346 &  7.7 &   63.5 &  9.6 & ... &  3.54 & <0.1        	     	  &   2.5$^{+0.4}_{-0.4}$ &  69.02 & 41.50 & L \\   	  
 587727884161581244 &  29.994183 &   -8.826961 &  0.392 & CXOMP J015958.6-084937 &  2.1 &   19.3 & 34.8 & ... &  2.11 & <2.1        	     	  &   1.3$^{+0.6}_{-0.6}$ &   7.32 & 42.35 & H {\sc ii} \\  
   587727884161581256 &  29.955580 &   -8.833209 &  0.405 & CXOMP J015949.3-084959 &  1.6 &  917.4 & 34.8 & I &  2.10 & <0.2        	     	  &   1.9$^{+0.1}_{-0.1}$ & 254.49 & 43.88 & no-class \\
 587727942420988082 & 144.690018 &    0.990713 &  0.171 & CXOMP J093845.6+005926 &  1.7 &   16.9 &  1.2 & ... &  4.22 & <0.4        	     	  &   1.4$^{+0.6}_{-0.6}$ & 114.19 & 42.76 & T \\   	  
   587728307491897590 & 168.966110 &    1.498703 &  0.352 & CXOMP J111551.8+012955 &  2.1 &  286.1 & 15.1 & I &  4.37 & <0.2        	     	  &   2.1$^{+0.3}_{-0.2}$ & 168.98 & 43.52 & T \\   	  
   587728905564258619 & 120.236748 &   36.056530 &  0.287 & CXOMP J080056.8+360323 &  1.3 &  496.6 & 35.4 & I &  4.95 &  0.1$^{+ 0.1}_{-0.1}$     &   2.0$^{+0.3}_{-0.3}$ &  90.87 & 43.10 & no-class \\ 
 587729157893456002 & 196.494354 &    3.956814 &  0.023 & CXOMP J130558.6+035724 &  3.0 &   55.1 &112.0 & ... &  2.08 & <0.1        	     	  &   1.3$^{+0.4}_{-0.3}$ &   4.26 & 39.56 & T \\   	  
 587729157893456055 & 196.563110 &    3.931599 &  0.110 & CXOMP J130615.1+035553 &  1.6 &   16.8 &112.0 & ... &  2.08 & <0.1        	     	  &   2.5$^{+1.0}_{-0.8}$ &   0.76 & 39.91 & T \\   	  
   587729158440419656 & 219.591049 &    3.670265 &  0.224 & CXOMP J143821.8+034012 &  3.6 &   37.0 & 56.2 & I &  2.61 & <0.4        	     	  &   1.9$^{+0.5}_{-0.5}$ &   7.36 & 41.74 & no-class \\
 587729158440419712 & 219.609650 &    3.649648 &  0.235 & CXOMP J143826.3+033858 &  2.3 &   19.6 & 56.2 & ... &  2.61 & <0.1        	     	  &   3.8$^{+1.0}_{-0.9}$ &   3.42 & 40.79 & P \\   	  
 587729388215337158 & 139.517227 &   51.687130 &  0.186 & CXOMP J091804.1+514113 &  6.2 &   23.4 & 18.7 & ... &  1.48 & 12.1$^{+ 7.3}_{-5.3}$     &  -0.6$^{+0.6}_{-0.7}$ &  43.52 & 42.41 & T \\   	  
 587729751132667997 & 247.652924 &   40.131489 &  0.077 & CXOMP J163036.7+400753 &  4.2 &   10.8 & 26.0 & ... &  0.88 & <0.1        	     	  &   3.4$^{+1.2}_{-1.0}$ &   3.53 & 39.86 & T \\   	  
 587729751132668135 & 247.629547 &   40.156597 &  0.077 & CXOMP J163031.0+400923 &  2.7 &   20.6 & 24.8 & ... &  0.88 & <0.1        	     	  &   3.4$^{+1.0}_{-0.9}$ &   3.56 & 39.83 & T \\   	  
   587729752213815608 & 260.041504 &   26.625124 &  0.159 & CXOMP J172009.9+263730 &  2.2 &  566.4 & 24.0 & I &  3.89 & <0.2        	     	  &   2.1$^{+0.3}_{-0.2}$ & 209.66 & 42.84 & T \\   	  
   587729753280741824 & 250.473267 &   40.029160 &  0.466 & CXOMP J164153.5+400144 &  2.5 &   23.4 & 45.3 & I &  1.02 & <0.3        	     	  &   2.3$^{+0.8}_{-0.6}$ &   4.84 & 42.24 & no-class \\ 
 587730816826671317 & 340.846375 &   -9.518315 &  0.144 & CXOMP J224323.1-093105 &  6.4 &   32.9 & 18.2 & ... &  4.33 &  3.0$^{+ 1.6}_{-1.1}$     &   0.1$^{+0.4}_{-0.5}$ &  45.66 & 42.25 & S \\   	  
 587731185129816192 & 359.235931 &   -0.986925 &  0.032 & CXOMP J235656.6-005912 &  1.2 &    9.2 & 17.9 & ... &  3.52 & <0.3        	     	  &   2.1$^{+1.3}_{-0.9}$ &   4.77 & 39.73 & L \\   	  
 587731186729550057 & 334.344727 &    0.351939 &  0.095 & CXOMP J221722.7+002106 &  6.1 &   14.6 & 75.4 & ... &  4.61 & <0.5         	    	  &   1.5$^{+0.8}_{-0.7}$ &   2.23 & 40.51 & T \\   	  
 587731187277430877 & 359.432159 &    0.654831 &  0.023 & CXOMP J235743.7+003917 &  7.4 &  147.9 & 11.9 & ... &  3.32 &  1.7$^{+ 0.6}_{-0.5}$     &   0.9$^{+0.2}_{-0.2}$ & 438.43 & 41.62 & S \\   	  
 587731187277430877 & 359.431946 &    0.655582 &  0.023 & CXOMP J235743.6+003920 &  7.4 &  142.7 & 11.9 & ... &  3.32 &  1.9$^{+ 0.7}_{-0.6}$     &   0.5$^{+0.3}_{-0.3}$ & 525.45 & 41.73 & S \\   	  
 587731187813253146 & 357.074768 &    1.104287 &  0.092 & CXOMP J234817.9+010615 &  7.7 &   36.5 & 48.7 & ... &  3.96 & <0.04        	     	  &   3.4$^{+0.7}_{-0.6}$ &   6.36 & 40.25 & P \\   	  
   587731511532453955 &  19.723360 &   -1.002003 &  0.045 & CXOMP J011853.6-010007 &  1.7 &  217.6 & 38.3 & I &  3.67 &  0.7$^{+ 0.4}_{-0.3}$     &   5.7$^{+2.0}_{-1.5}$ &  28.49 & 40.62 & no-class \\ 
 587731512613404741 &  36.155399 &   -0.047232 &  0.127 & CXOMP J022437.2-000250 &  0.8 &   12.0 & 86.7 & ... &  2.85 & <0.6        	     	  &   1.7$^{+0.9}_{-0.7}$ &   1.21 & 40.47 & T \\   	  
 587731513148571800 &  32.354515 &    0.399077 &  0.061 & CXOMP J020925.0+002356 &  9.9 &   29.2 &  2.6 & ... &  2.79 & <0.4        	     	  &   1.6$^{+0.5}_{-0.5}$ & 138.82 & 41.88 & T \\   	  
 587731868557377700 & 168.673538 &   53.250507 &  0.106 & CXOMP J111441.6+531501 &  4.8 &   23.4 & 16.8 & ... &  0.92 &  7.2$^{+ 5.3}_{-2.7}$     &  -0.6$^{+0.6}_{-0.6}$ &  43.22 & 41.96 & L \\   	  
 587731873388101721 & 123.742249 &   36.890556 &  0.108 & CXOMP J081458.1+365326 &  8.2 &   10.6 &  9.3 & ... &  5.02 & <0.2       	     	  &   2.4$^{+1.5}_{-1.1}$ &  10.99 & 41.09 & S \\   	  
 587731886272741666 & 123.901619 &   36.765156 &  0.174 & CXOMP J081536.3+364554 &  6.1 &    9.3 &  8.8 & ... &  5.00 & <0.3      	     	  &   3.1$^{+1.3}_{-1.0}$ &  10.98 & 41.30 & P \\   	  
   587732050021318723 & 148.205582 &   51.884888 &  0.215 & CXOMP J095249.3+515305 &  1.1 &  283.2 & 24.4 & I &  0.88 & <0.5      	     	  &   2.6$^{+0.6}_{-0.5}$ &  59.21 & 42.60 & L \\   	  
 587732135913521261 & 150.952209 &   47.631630 &  0.051 & CXOMP J100348.5+473753 &  0.9 &   13.5 & 13.5 & ... &  0.93 & <0.1        	     	  &   2.7$^{+1.0}_{-0.8}$ &   4.57 & 39.90 & T \\   	  
 587732469851553889 & 131.364151 &   34.419083 &  0.025 & CXOMP J084527.3+342508 &  4.3 &    4.6 &  4.0 & ... &  3.42 & <0.3        	     	  &   2.6$^{+2.2}_{-1.4}$ &   7.92 & 39.56 & T \\   	  
 587732482200371401 & 145.504807 &   41.440948 &  0.243 & CXOMP J094201.1+412627 &  6.3 &   11.4 &  7.5 & ... &  0.95 & <0.1        	     	  &   4.0$^{+1.5}_{-1.2}$ &   6.62 & 41.01 & P \\   	  
 587732484351983843 & 156.306671 &   47.115517 &  0.060 & CXOMP J102513.6+470655 & 10.0 &    5.8 &  2.1 & ... &  1.22 & <1.0        	     	  &   1.9$^{+1.3}_{-1.1}$ &  31.99 & 41.16 & T \\   	  
   587733397572747445 & 244.418823 &   35.004284 &  0.029 & CXOMP J161740.5+350015 &  0.6 &  151.7 & 18.3 & I &  1.47 & <0.02       	     	  &   2.7$^{+0.4}_{-0.4}$ &  35.71 & 40.29 & L \\   	  
 587733603191161025 & 240.295242 &   43.194328 &  0.071 & CXOMP J160110.8+431139 &  7.5 &  255.5 & 26.7 & ... &  1.31 & <5.8        	     	  &   0.9$^{+0.9}_{-0.8}$ & 294.80 & 42.47 & S \\   	  
 587733604801708195 & 241.492157 &   44.055454 &  0.044 & CXOMP J160558.1+440319 &  3.6 & 1324.3 &  4.6 & ... &  1.15 & <0.01       	     	  &   1.9$^{+0.0}_{-0.0}$ &1619.01 & 42.56 & S \\   	  
 587734304342212857 & 336.311615 &   -0.364847 &  0.142 & CXOMP J222514.7-002153 &  7.5 &    6.2 &  3.1 & ... &  5.12 & <0.5        	     	  &   2.5$^{+1.5}_{-1.2}$ &  29.36 & 41.76 & T \\   	  
 587735348561051857 & 151.805237 &   12.787229 &  0.248 & CXOMP J100713.2+124714 &  4.0 &   18.0 & 36.3 & ... &  3.68 & <0.1        	     	  &   4.2$^{+1.1}_{-1.0}$ &   2.60 & 40.55 & no-class \\   
   587735661546504321 & 143.357269 &   34.048050 &  0.027 & CXOMP J093325.7+340252 &  2.5 &   33.6 & 33.6 & I &  1.47 & <0.03       	     	  &   3.6$^{+0.6}_{-0.6}$ &   9.81 & 39.24 & L \\   	  
 587735661546504414 & 143.272659 &   34.062729 &  0.277 & CXOMP J093305.4+340345 &  6.8 &   23.1 & 33.6 & ... &  1.47 & <0.4        	     	  &   1.5$^{+0.6}_{-0.6}$ &   8.69 & 42.08 & T \\   	  
 587735662089666625 & 159.883911 &   39.836349 &  0.068 & CXOMP J103932.1+395010 &  2.7 &   10.7 &  5.0 & ... &  1.43 & <0.9        	     	  &   0.8$^{+0.7}_{-0.7}$ &  25.60 & 41.33 & T \\   	  
 587735666377883763 & 206.114655 &   56.024902 &  0.070 & CXOMP J134427.5+560129 &  7.5 &   45.1 & 43.5 & ... &  1.09 &  0.9$^{+ 0.6}_{-0.4}$     &   0.8$^{+0.4}_{-0.4}$ &  16.90 & 41.19 & no-class \\    
 587735666377883859 & 206.093246 &   55.951073 &  0.038 & CXOMP J134422.3+555703 &  3.9 &    7.4 & 42.7 & ... &  1.09 &  <0.2      	  	  &   3.2$^{+1.3}_{-1.1}$ &   1.22 & 38.85 & T \\   	  
 587735695377432678 & 214.062256 &   53.146236 &  0.114 & CXOMP J141614.9+530846 &  5.7 &    8.1 & 60.5 & ... &  1.27 &  <22.5     	   	  &  -0.1$^{+0.9}_{-1.1}$ &   3.95 & 40.99 & no-class \\    
 587735696440623133 & 171.764420 &   56.902294 &  0.055 & CXOMP J112703.4+565408 &  3.3 &    7.8 & 38.3 & ... &  0.92 &  <0.5       	  	  &   1.5$^{+1.3}_{-1.0}$ &   2.26 & 40.02 & no-class \\    
 587735696440623158 & 171.882370 &   56.876919 &  0.005 & CXOMP J112731.7+565236 &  0.8 &   45.3 & 39.3 & ... &  0.90 &  <0.4       	  	 &  0.6$^{+0.4}_{-0.4}$ &  16.00 & 38.99 & H {\sc ii} \\  
 587735696440623212 & 171.780807 &   56.828369 &  0.175 & CXOMP J112707.3+564942 &  3.8 &   12.0 & 38.3 & ... &  0.91 &  <0.1      	  	 &   3.8$^{+1.6}_{-1.1}$ &   2.48 & 40.32 & P \\   	  
 587735696448880650 & 206.089798 &   55.856201 &  0.037 & CXOMP J134421.5+555122 &  4.1 &    9.6 & 42.7 & ... &  1.08 &  <0.2      	  	 &   2.2$^{+1.2}_{-0.9}$ &   1.70 & 39.35 & no-class \\    
 587735696451371138 & 215.187088 &   53.653847 &  0.117 & CXOMP J142044.9+533913 &  6.4 &  126.2 &  4.9 & ... &  1.19 &  <0.1      	  	 &   1.9$^{+0.4}_{-0.3}$ & 167.86 & 42.48 & S \\   	  
 587735696979656747 & 180.807953 &   57.890541 &  0.034 & CXOMP J120313.9+575325 &  5.9 &   93.9 & 57.8 & ... &  1.43 &  <0.2      	  	 &   1.9$^{+0.3}_{-0.2}$ &  19.12 & 40.43 & H {\sc ii} \\  
 587735744228753459 & 245.066696 &   29.488642 &  0.061 & CXOMP J162016.0+292919 &  0.6 &   10.7 & 32.2 & ... &  2.74 &  <12.6      	  	 &   0.1$^{+0.7}_{-0.8}$ &   6.82 & 40.69 & no-class \\    
   587736584961196166 & 205.136429 &   40.293945 &  0.171 & CXOMP J134032.7+401738 &  1.2 &   47.2 & 43.1 & I &  0.80 &  <0.1      	  	 &   2.8$^{+0.5}_{-0.4}$ &   5.02 & 41.03 & P \\   	  
 587736752468394214 & 241.406998 &   32.936256 &  0.053 & CXOMP J160537.6+325610 &  7.7 &   18.7 & 19.9 & ... &  2.30 &  <0.2      	  	 &   1.9$^{+0.8}_{-0.7}$ &  10.33 & 40.53 & T \\   	  
 587736752468394267 & 241.437714 &   32.872177 &  0.115 & CXOMP J160545.0+325219 &  6.0 &   30.3 & 20.1 & ... &  2.31 &  <0.1      	  	 &   2.1$^{+0.5}_{-0.5}$ &  16.30 & 41.41 & no-class \\    
 587736781993935128 & 236.260956 &   36.156464 &  0.060 & CXOMP J154502.6+360923 &  3.2 &    9.8 & 18.8 & ... &  1.65 &  <0.1      	  	 &   3.3$^{+1.2}_{-1.0}$ &   2.61 & 39.55 & no-class \\    
 587736781993935188 & 236.282913 &   36.146473 &  0.069 & CXOMP J154507.8+360847 &  3.5 &   12.5 & 18.8 & ... &  1.65 &  <0.1       	  	 &   3.6$^{+1.3}_{-1.1}$ &   2.91 & 39.54 & no-class \\    
 587736941445447871 & 216.464218 &   35.567928 &  0.186 & CXOMP J142551.4+353404 &  3.2 &    7.8 & 50.3 & ... &  1.10 &  <0.2       	  	 &   3.5$^{+1.8}_{-1.6}$ &   1.20 & 40.22 & S \\   	  
   588007004192637004 & 243.921677 &   47.186592 &  0.198 & CXOMP J161541.2+471111 &  0.5 &  121.6 &  3.4 & I &  1.23 &  <0.1       	  	 &   1.7$^{+0.4}_{-0.3}$ & 239.03 & 43.18 & T \\   	  
 588007004192637199 & 243.922241 &   47.167870 &  0.197 & CXOMP J161541.3+471004 &  1.4 &    5.8 &  3.4 & ... &  1.23 &  <0.1       	  	 &   3.6$^{+3.9}_{-1.6}$ &   6.97 & 40.99 & P \\   	  
 588007004694839559 & 119.126747 &   41.036259 &  0.072 & CXOMP J075630.4+410210 &  3.3 &  109.5 &  6.2 & ... &  4.48 &  <0.04       	  	 &   2.5$^{+0.5}_{-0.4}$ & 102.97 & 41.65 & L \\   	  
 588007005769957640 & 120.440094 &   44.110588 &  0.131 & CXOMP J080145.6+440638 &  6.6 &   11.8 &  9.2 & ... &  4.73 &  <0.1       	  	 &   3.5$^{+1.2}_{-1.0}$ &   7.35 & 40.67 & P \\   	  
 588010879292932251 & 173.027710 &    4.893930 &  0.150 & CXOMP J113206.6+045338 &  3.4 &    6.5 &  6.3 & ... &  3.43 &  <0.2       	  	 &   3.2$^{+2.4}_{-1.4}$ &   4.43 & 40.68 & no-class \\    
 588016891170914329 & 143.504242 &   33.990730 &  0.027 & CXOMP J093401.0+335926 &  5.9 &   13.6 & 31.8 & ... &  1.47 &  <0.1        	  	 &   3.4$^{+1.2}_{-1.0}$ &   4.19 & 38.94 & no-class \\    
 588017111297622182 & 189.892838 &   47.537556 &  0.131 & CXOMP J123934.2+473215 &  9.1 &   82.4 &  4.5 & ... &  1.13 &   2.6$^{+ 0.7}_{-0.6}$   &   0.2$^{+0.2}_{-0.3}$ & 434.23 & 43.15 & S \\   	  
 588017111298932927 & 194.301773 &   47.330669 &  0.131 & CXOMP J125712.4+471950 &  2.3 &   31.3 & 49.8 & ... &  1.15 &  <0.1          	  	 &   2.2$^{+0.6}_{-0.5}$ &   3.13 & 40.78 & no-class \\    
 588017111833968667 & 187.945435 &   47.927235 &  0.030 & CXOMP J123146.9+475538 &  5.5 &    9.5 &  6.5 & ... &  1.17 &  <0.7          	  	 &  -0.2$^{+0.9}_{-1.2}$ &  34.97 & 40.81 & S \\   	  
 588017567628591240 & 164.638321 &   12.720986 &  0.119 & CXOMP J105833.1+124315 &  6.2 &    8.8 &  4.9 & ... &  2.13 &  <0.5          	  	 &   1.9$^{+1.0}_{-0.9}$ &  29.98 & 41.76 & T \\   	  
 588017569779023967 & 201.370422 &   11.335633 &  0.086 & CXOMP J132528.9+112008 &  5.5 &   75.7 &  4.4 & ... &  1.92 &  <0.2          	  	 &   1.9$^{+0.3}_{-0.3}$ & 124.39 & 42.06 & S \\   	  
 588017604148133957 & 186.976318 &   44.363770 &  0.276 & CXOMP J122754.3+442149 &  1.7 &   21.6 &  4.8 & ... &  1.34 &  <1.3          	  	 &   1.4$^{+0.5}_{-0.5}$ &  48.44 & 42.83 & T \\   	  
 588017605772837051 & 228.612808 &   36.634247 &  0.161 & CXOMP J151427.0+363803 &  2.5 &   21.0 & 43.5 & ... &  1.35 &  <0.2          	  	 &   2.4$^{+0.8}_{-0.7}$ &   2.28 & 40.80 & P \\   	  
 588017625613795424 & 170.110840 &   43.255371 &  0.145 & CXOMP J112026.6+431519 &  7.0 &    1.8 & 17.6 & ... &  1.96 &  <39.4         	  	 &  -0.7$^{+2.1}_{-3.0}$ &   3.62 & 41.14 & S \\   	  
 588017720102813842 & 168.741943 &   40.603119 &  0.075 & CXOMP J111458.0+403611 &  4.9 &   22.1 & 28.3 & ... &  1.94 &  <0.8          	  	 &   0.7$^{+0.5}_{-0.5}$ &  10.14 & 41.03 & S \\   	  
 588017722259865662 & 198.026779 &   42.690838 &  0.179 & CXOMP J131206.4+424127 &  3.2 &   15.4 & 80.6 & ... &  1.37 &  <2.2          	  	 &  -0.3$^{+0.7}_{-1.0}$ &   4.76 & 41.44 & no-class \\    
 588017722259865691 & 198.164169 &   42.713638 &  0.111 & CXOMP J131239.3+424248 &  3.0 &   64.5 & 90.7 & ... &  1.37 &  <0.3          	  	 &   1.6$^{+0.3}_{-0.3}$ &   6.32 & 41.08 & L \\   	  
 588017947748139147 & 205.227905 &   40.109875 &  0.170 & CXOMP J134054.6+400635 & 10.5 &  917.5 & 45.1 & ... &  0.79 &   0.04$^{+ 0.0}_{-0.0}$  &   2.0$^{+0.2}_{-0.1}$ & 154.08 & 42.80 & L \\   	  
 588018089466658843 & 231.747757 &   35.976982 &  0.055 & CXOMP J152659.4+355837 &  0.6 &   22.4 &  9.1 &   I &  1.56 &  <0.1          	  	 &   2.8$^{+0.7}_{-0.6}$ &  11.71 & 40.35 & T \\   	  
 588295840714129492 & 185.380676 &   49.176918 &  0.184 & CXOMP J122131.3+491036 &  4.4 &   17.2 & 72.2 & ... &  1.42 &  <0.1       	  	 &   3.1$^{+1.0}_{-0.9}$ &   1.97 & 40.60 & T \\   	  
 588295840714129502 & 185.415253 &   49.332184 &  0.124 & CXOMP J122139.6+491955 &  5.0 &   15.3 & 77.0 & ... &  1.44 &  <3.1          	  	 &   0.5$^{+0.8}_{-0.8}$ &   4.16 & 41.09 & T \\   	  
 588295842853224515 & 156.433044 &   47.326378 &  0.062 & CXOMP J102543.9+471934 &  5.7 &    2.9 &  1.9 & ... &  1.26 &  <0.3          	  	 &   3.7$^{+2.6}_{-2.0}$ &  12.03 & 40.00 & P \\   	  
 588297863638089749 & 126.979630 &   29.449829 &  0.029 & CXOMP J082755.1+292659 &  9.8 &   15.0 & 14.6 & ... &  3.73 &  <0.2          	  	 &   2.3$^{+1.3}_{-0.9}$ &   8.64 & 39.84 & H {\sc ii} \\  
 588848898846752926 & 197.902756 &   -0.922150 &  0.083 & CXOMP J131136.6-005519 &  8.6 &  100.3 & 20.1 & ... &  1.78 &   7.3$^{+ 5.2}_{-3.9}$   &   0.2$^{+0.7}_{-0.7}$ & 162.69 & 42.34 & S \\   	  
 588848898846818469 & 198.065002 &   -0.930759 &  0.081 & CXOMP J131215.6-005550 &  4.0 &    3.6 & 20.7 & ... &  1.79 &  <0.2          	  	 &   4.2$^{+2.8}_{-1.8}$ &   2.26 & 39.28 & P \\   	  
 588848899377529043 & 183.972015 &   -0.601967 &  0.119 & CXOMP J121553.2-003607 &  3.0 &    8.1 & 42.0 & ... &  2.08 &   0.6$^{+ 1.0}_{-0.6}$   &   1.4$^{+0.9}_{-0.9}$ &   2.38 & 40.75 & L \\   	  
 588848899377529068 & 184.006836 &   -0.625664 &  0.121 & CXOMP J121601.6-003732 &  4.9 &   17.4 & 41.6 & ... &  2.09 &  <0.0          	  	 &   4.9$^{+1.1}_{-1.0}$ &   5.90 & 39.71 & P \\   	  
 588848899387228268 & 206.030533 &   -0.475767 &  0.101 & CXOMP J134407.3-002832 &  0.5 &    3.6 &  8.7 & ... &  2.02 &  <2.5          	  	 &   1.2$^{+1.3}_{-1.2}$ &   3.75 & 40.82 & T \\   	  
 588848901519179802 & 170.660416 &    1.059925 &  0.075 & CXOMP J112238.4+010335 &  3.4 &   69.2 & 18.5 & ... &  4.10 &  <0.1          	  	 &   1.8$^{+0.3}_{-0.3}$ &  39.08 & 41.46 & T \\   	  
 588848901519179967 & 170.649841 &    1.116308 &  0.039 & CXOMP J112235.9+010658 &  3.4 &    6.3 & 19.9 & ... &  4.11 &  <0.5          	  	 &   2.5$^{+1.3}_{-1.1}$ &   3.05 & 39.58 & H {\sc ii} \\  
 588848901528289406 & 191.438202 &    1.079846 &  0.106 & CXOMP J124545.1+010447 &  1.1 &    9.7 &  6.3 & ... &  1.69 &  <0.1          	  	 &   3.5$^{+1.3}_{-1.1}$ &   6.31 & 40.35 & T \\   	  
 588848901531107518 & 197.878723 &    1.184934 &  0.070 & CXOMP J131130.8+011105 &  8.5 &   24.7 &  4.5 & ... &  1.97 &  15.1$^{+ 8.5}_{-5.8}$   &  -1.5$^{+0.7}_{-0.8}$ & 270.20 & 42.40 & S \\           
\enddata
\tablecomments{(1) SDSS Object ID. (2) and (3) J2000 epoch. (4)
  spectroscopic redshift. (5) ChaMP X-ray source ID, expressed as
  CXOMP Jhhmmss.s+/-ddmmss, using the truncated X-ray source
  position. (6) $Chandra$ off-axis angle in arcmin; (7) net 0.5 -- 8
  keV source counts. (8) Vignetting-corrected exposure time in
  ksec. (9) I = intended $Chandra$ PI target.  (10) Galactic column in
  units of $10^{20}$ cm$^{-2}$. (11) Best-fit YAXX intrinsic column
  density in $10^{22}$ cm$^{-2}$; upper limits are at 90\% confidence
  level and errors represent 1-$\sigma$ uncertainties. (12) Best-fit
  YAXX power-law index $\Gamma$; errors represent 1-$\sigma$
  uncertainties. (13) X-ray flux (0.5 -- 8 keV) in units of 10$^{-15}$ erg s$^{-1}$
  cm$^{-2}$. (14) log X-ray luminosity (0.5 -- 8 keV) in erg
  s$^{-1}$. (15) Optical spectral classification, S = Seyfert, L =
  LINER, T = Transition Object, P = Passive galaxy; see text for
  details.  Note that only for objects with > 200 counts, $N_H^{Intr.}$ and
  $\Gamma$ values are the result of fitting simultaneously a power-law
  and absorption. }
\end{deluxetable}

\clearpage
\end{landscape}

\begin{deluxetable}{lccc}
\tablecolumns{4} \tablewidth{0pt}
\tablecaption{Object Sample Statistics.
\label{stats}}
\tablehead{
\colhead{Sample} &
\colhead{$N_{\rm opt}$(fraction\tablenotemark{a},\%)} &
\colhead{$N_{\rm X-ray~det.}$(fraction\tablenotemark{b},\%)} &
\colhead{$N_{\rm X-ray~det.}$/$N_{\rm opt}$(\%)}}
\startdata
H II                  &498 (27.6)  &  7 (6.5)&  1.4 $\pm$ 4.5\\
Seyfert               & 28 ({\bf 1.5})&18 ({\bf 16.8})&{\bf 64.3 $\pm$ 18.9}\\
Transition            &194 (10.7)  & 32 (29.9)& 16.5 $\pm$ 7.2\\
LINER                 & 70 ( 3.9)  & 13 (12.1)& 18.6 $\pm$ 11.9\\
some emision--no class&558 (31.1)  & 19 (17.8)&  3.4 $\pm$ 4.2\\ 
Passive               &459 (25.4)  & 18 (16.8)&  3.9 $\pm$ 4.6\\
\enddata
\tablenotetext{a}{Fraction by type of all 1807 SDSS galaxies on ACIS
  chips, which excludes those falling on $ccd=8$, and with $\theta >
  0.2$ deg.}   
\tablenotetext{b}{Fraction of all 107 X-ray detected SDSS galaxies.}   
\tablecomments{The quoted errors represent standard deviations
  assuming Poisson statistics.}
\end{deluxetable}

\begin{deluxetable}{lccc}
\tablecolumns{4} \tablewidth{0pt}
\tablecaption{ $\Gamma - L/L_{\rm edd}$ - Correlation coefficients and Significance 
\label{spearman}}
\tablehead{
\colhead{Sample} &
\colhead{$r_s$} &
\colhead{Prob.} &
\colhead{N}}
\startdata
\sidehead{ for $L_X = L_X (0.5 - 8)$ keV .........}
 all       & {\bf -0.52}  & $< 1.0 \times 10^{-3}$& 107 \\
Seyfert    &      -0.18   & $ 0.43$               &  18\\
Transition &      -0.29  & $ 0.11$               &  31\\
LINER      &      -0.39   & $ 0.18$               &  13\\
\sidehead{ for $L_X = L_X (2 - 10)$ keV .........}
 all       & {\bf -0.75}  & $< 1.0 \times 10^{-3}$& 107 \\
Seyfert    &      -0.36   & $ 0.12$               &  18\\
Transition & {\bf -0.52}  & $  2.3 \times 10^{-3}$&  31\\
LINER      & {\bf -0.56}  & $  4.4 \times 10^{-2}$&  13\\
\enddata
\tablecomments{The Spearman-rank test correlation coefficient, chance
  probability, and number of sources for each correlation/sample
  respectively.  The most significant probabilities for anticorrelations (chance
probability less than 5\%) are shown in boldface. }
\end{deluxetable}

\begin{deluxetable}{lccc}
\tablecolumns{4} \tablewidth{0pt}
\tablecaption{$\Gamma - L/L_{\rm edd}$ - Linear Regression Coefficients
\label{lsf}}
\tablehead{
\colhead{Sample} &
\colhead{Slope} &
\colhead{Intercept} &
\colhead{$\chi^2$/dof}}
\startdata
\sidehead{ for $L_X = L_X (0.5 - 8)$ keV .........}
all       & -0.16 $\pm$ 0.04 & 1.36 $\pm$ 0.13 & 309/105\\
all, no passive, no H{\sc ii}          &-0.07 $\pm$ 0.05 & 1.51 $\pm$ 0.15 & 235/81\\
all with $L_X \ga 10^{42}$ erg s$^{-1}$& 0.12 $\pm$ 0.12 & 2.09 $\pm$ 0.29 & 83/17\\
Seyfert    &  0.94 $\pm$ 0.15 & 3.59 $\pm$ 0.35 & 89/16\\
Transition & -0.001 $\pm$ 0.11 & 1.88 $\pm$ 0.37 & 21/29\\
LINER      & -0.20 $\pm$ 0.09 & 1.42 $\pm$ 0.34 & 29/11\\

\sidehead{ for $L_X = L_X (2 - 10)$ keV ..........}
all       & -0.27 $\pm$ 0.04 & 0.98 $\pm$ 0.13 & 274/105\\
all, no passive, no H{\sc ii}          &-0.21 $\pm$ 0.05& 1.11 $\pm$ 0.15& 219/81\\
all with $L_X \ga 10^{42}$ erg s$^{-1}$&-0.19 $\pm$ 0.14& 1.28 $\pm$ 0.36& 95/16\\
Seyfert    &  0.42 $\pm$ 0.18 & 2.49 $\pm$ 0.43 & 119/16\\
Transition & -0.09 $\pm$ 0.10 & 1.57 $\pm$ 0.38 & 21/29\\
LINER      & -0.25 $\pm$ 0.09 & 1.18 $\pm$ 0.34 & 26/11\\
\enddata
\tablecomments{The slope, intercept, $\chi^s$,
  and degrees of freedom (dof) for the best error-weighted
  linear fits for the whole sample of X-ray detected galaxies, and per
  spectral type.  Errors are at 1-$\sigma$ confidence levels.}
\end{deluxetable}

\end{document}